\documentclass[pra,twocolumn,showpacs,groupedaddress,notitlepage,superscriptaddress,floatfix,nofootinbib]{revtex4-2}

\usepackage[english]{babel}
\usepackage{amsmath}
\usepackage{amssymb}
\usepackage{amsthm}
\usepackage{amsfonts}
\usepackage{braket}%Dirac Notation in QM
\usepackage{cases}
\usepackage{color}
\usepackage{xcolor}
\usepackage{graphicx,subfigure}% hy77 Include figure files
\usepackage{multirow}
\usepackage{float}
\usepackage{bm}% bold math
\usepackage{hyperref}
\hypersetup{
    colorlinks=true, 
    linkcolor=cyan,
    citecolor=magenta, 
    filecolor=magenta, 
    urlcolor=cyan,
    runcolor=cyan
}
\usepackage[capitalise]{cleveref}
\usepackage{booktabs} 

\newtheorem{mytheorem}{Theorem}
\newtheorem{mylemma}{Lemma}

\newtheorem{mydefinition}{Definition}

\newcommand{\ignore}[1]{}

\newcommand{\abs}[1]{\left|#1\right|}
\newcommand{\Ugrp}{\mathsf{U}}
\newcommand{\ave}[1]{\<{#1}\>}

\usepackage{commath}
\usepackage{color}
\usepackage{amsmath}
\usepackage{adjustbox}
\usepackage[normalem]{ulem}

\newcommand{\bes} {\begin{subequations}}
\newcommand{\ees} {\end{subequations}}
\newcommand{\beq}{\begin{equation}}
\newcommand{\eeq}{\end{equation}}

\def\ox{\otimes}
\def\>{\rangle}
\def\<{\langle}

\def\XY4{\text{XY}4}

\def\mC{\mathcal{C}}
\def\mD{\mathcal{D}}
\def\mG{\mathcal{G}}
\def\mH{\mathcal{H}}
\def\mL{\mathcal{L}}

\def\mP{\mathcal{P}}

\def\mS{\mathcal{S}}
\def\NS{\mathcal{N}(\mS)}
\def\tNS{\tilde{\mathcal{N}}(\mS)}
\def\tmG{\tilde{\mathcal{G}}}
\def\tmL{\tilde{\mathcal{L}}}
\def\tmS{\tilde{\mathcal{S}}}
\def\tg{\tilde{g}}

\def\HSBstab{H_{SB}^{\mS_\star}}
\def\HSBlog{H_{SB}^{\mL_\star}}
\def\HSBdet{H_{SB}^{\mD_\star}}

\begin{document}

\title{Demonstration of High-Fidelity Entangled Logical Qubits 
using Transmons}

\author{Arian Vezvaee}
\affiliation{Department of Electrical \& Computer  Engineering, University of Southern California, Los Angeles, CA 90089}
\affiliation{Center for Quantum Information Science \& Technology, University of
Southern California, Los Angeles, CA 90089}
\affiliation{Quantum Elements, Inc., Thousand Oaks, CA}

\author{Vinay Tripathi}
\thanks{current affiliation: IBM Quantum}
\affiliation{Center for Quantum Information Science \& Technology, University of
Southern California, Los Angeles, CA 90089}
\affiliation{Department of Physics \& Astronomy, University of Southern California,
Los Angeles, CA 90089}

\author{Mario Morford-Oberst}
\affiliation{Department of Electrical \& Computer  Engineering, University of Southern California, Los Angeles, CA 90089}
\affiliation{Center for Quantum Information Science \& Technology, University of
Southern California, Los Angeles, CA 90089}

\author{Friederike Butt}
\affiliation{Institute for Quantum Information, RWTH Aachen University, Aachen, Germany}
\affiliation{Institute for Theoretical Nanoelectronics (PGI-2), Forschungszentrum J\"{u}lich, J\"{u}lich, Germany}

\author{Victor Kasatkin}
\affiliation{Department of Electrical \& Computer  Engineering, University of Southern California, Los Angeles, CA 90089}
\affiliation{Center for Quantum Information Science \& Technology, University of
Southern California, Los Angeles, CA 90089}

\author{Daniel A. Lidar}
\affiliation{Department of Electrical \& Computer  Engineering,  University of Southern California, Los Angeles, 
CA 90089}
\affiliation{Center for Quantum Information Science \& Technology, University of
Southern California, Los Angeles, CA 90089}
\affiliation{Quantum Elements, Inc., Thousand Oaks, CA}
\affiliation{Department of Physics \& Astronomy, University of Southern California,
Los Angeles, CA 90089}
\affiliation{Department of Chemistry, University of Southern California,
Los Angeles, CA 90089}

\begin{abstract}
Quantum error correction (QEC) codes are necessary to fault-tolerantly operate quantum computers. However, every such code is inherently limited by its inability to detect logical errors. Here, we propose and implement a method that leverages dynamical decoupling (DD) to drastically suppress logical errors. The key to achieving this is to use the 
{normalizer elements} of the QEC code as DD pulses, which we refer to as {normalizer}
dynamical decoupling ({N}DD). The resulting hybrid QEC-{N}DD strategy is in principle capable of handling arbitrary weight errors.
We test QEC-{N}DD using IBM transmon devices and the $[[4,2,2]]$ code, demonstrating performance that significantly exceeds the capabilities of using either this code or DD in isolation. We present a method that allows for the detection of logical errors affecting logically encoded Bell states, which, in this case, arise primarily from crosstalk among physical qubits. Building on this, we experimentally demonstrate high-fidelity entangled logical qubits. The fidelities we achieve are beyond-breakeven, i.e., they significantly exceed the corresponding fidelities of unprotected entangled qubits in the same setting.
\end{abstract}

\maketitle

\section{Introduction}

Quantum error correction (QEC)~\cite{Shor1995PRA,Steane:96a,Gottesman1996PRA,Calderbank1996} is fundamental to the realization of fault-tolerant quantum computation~\cite{DiVincenzo:96,Aharonov:96,Gottesman:97a,Knill:98}, ensuring the preservation of quantum information undergoing error processes during computation and storage~\cite{LidarBrun2013QEC,Campbell:2017aa}. Numerous successful experimental demonstrations of QEC across various platforms have been reported over the years~\cite{PhysRevLett.81.2152,chiaverini_realization_2004,Pittman:05b,Takita2017,Harper:2019aa}, with the scale and pace accelerating recently towards genuine fault tolerance~\cite{Linke:aa, Krinner2022Nature,GoogleAI2023QEC,Bluvstein2023Nature, ButtPRX2024,GoogleAI2024Nature,reichardt2024demonstrationquantumcomputationerror, GoogleAI2024NatureColorCode, GoogleAI2024arxivDynamicSurfaceCode}. 

Fault-tolerant quantum computing will require the stability of logical qubits on the long timescales of quantum algorithms that solve utility-scale problems~\cite{Reiher:2017aa,Gidney2021Quantum,AWS2023arxiv}. Above threshold, on such timescales, low-weight physical errors may transform into logical errors, i.e., errors of weight higher than a deployed fixed-distance QEC code can handle.
This can become a concern even below threshold, as long-range spatial and temporal correlations may develop that violate the assumptions underlying fault tolerance theory~\cite{PhysRevA.73.052311,preskill:12,Clader:2021aa,sriram2024nonuniformnoiseratesgriffiths,kam2024detrimentalnonmarkovianerrorssurface}.

Conventionally, suppressing higher-weight errors requires increasing the QEC code distance, e.g., by means of code concatenation~\cite{Knill:96a,Aliferis:05}, or increasingly larger codes such as is done with surface codes~\cite{Dennis:02,Fowler:2012ys}, color codes~\cite{Bombin:2006aa}, or quantum low-density parity-check codes~\cite{Breuckmann:2021aa}. Although effective, these strategies result in significant overhead, substantially increasing the number of physical qubits required and the time required to decode and correct errors. 

Here, we propose and demonstrate a low-cost method that combines a \emph{fixed-distance} QEC or quantum error detection (QED) code with dynamical decoupling (DD)~\cite{Viola:98,Viola1999PRL,Duan:98e,Vitali:99}. This hybrid method, which we call QEC-{N}DD, can handle \emph{arbitrary-weight} errors. 
Here, {N}DD stands for ``{normalizer} dynamical decoupling''; the decoupling sequence is implemented using the {normalizer elements} of the QEC code as pulses. 

Standard, ``physical'' DD, where DD pulses act not at the logical level but at the physical qubit level, has recently shown great progress, improving the fidelity of storing quantum states~\cite{Tripathi2025PRL, Pokharel2018PRL,Souza2021QIP,Ezzell2021PRApp,tong2024arxiv,seif2024arxiv,rahman2024arxiv}, circuits~\cite{GoogleAI2019Sup,Jurcevic2021IOP,baumer2023arxiv}, and even the performance of entire algorithms~\cite{Pokharel2023PRL,Pokharel2024npj,singkanipa2024demonstration,Baumer2024}. Furthermore, physical DD can be seamlessly combined with fault-tolerant QEC~\cite{Ng2011PRA,PazSilva2013SciRep} and several recent QEC experiments have used physical DD profitably \cite{GoogleAI2023QEC, GoogleAI2024Nature,Bluvstein2023Nature,Goto2023PRR,ButtPRX2024,han2024arxiv}. 
However, such pulse sequences can introduce additional errors due to control imperfections in the pulses and due to crosstalk, potentially overshadowing the benefits of DD. To address this challenge, we design our {N}DD sequences to be robust against such control errors~\cite{Quiroz2013PRA,Genov2017PRL} and crosstalk~\cite{Wocjan:2006aa,Tripathi2022PRApp,Zhou2023PRL,evert2024syncopated,brown2024efficient}, ensuring that the advantages of logical error suppression are not compromised. This robust design allows our {N}DD implementation to enhance the protection of the code space as intended, effectively targeting both logical and control errors. We demonstrate the practical utility and advantages of this approach using $127$-qubit IBM quantum processors.

The workhorse in our demonstrations is the $[[4,2,2]]$ quantum error detection code, whose two logical qubits we use to prepare logically encoded Bell states. To demonstrate the effectiveness of NDD in suppressing logical errors, we first need a method to detect such errors. However, this code is constrained by its low distance and is incapable of detecting prevalent logical errors, including $ZZ$ crosstalk errors. To overcome this limitation, we design experiments in which, through the use of logically encoded Bell states, we nevertheless unequivocally detect the occurrence of logical errors for a known input logical state. 
This is then followed by the implementation of various versions of {N}DD, each employing different subsets of 
{normalizer elements} to construct the sequence. These implementations effectively demonstrate significant suppression of logical errors and substantial improvement in the fidelity of the code space. 

The structure of this paper is as follows. In \cref{sec:background} we provide pertinent background on the error model and DD from the perspective of group averaging. We discuss standard Pauli group DD, as well as encoded DD, i.e., DD implemented using the stabilizers and/or logical operators of a stabilizer code. 
In \cref{sec:expt-design} we describe our experimental design and methodology.
We then turn to our experimental demonstration of high-fidelity logical Bell states in \cref{sec:expt-results}. This section describes our experimental design and methodology, evidence that both physical DD and error detection with postselection improve logical Bell state fidelity, and finally, the evidence that the hybrid QEC-{N}DD strategy significantly outperforms both standalone dynamical decoupling and error detection.

We conclude in \cref{sec:discussion} and provide supporting technical details in the Appendix.

%%%%%%%%%%%%%%%%%%%%%%%%%%%%
%%%%%%%%%%%%%%%%%%%%%%%%%%%%
%%%%%%%%%%%%%%%%%%%%%%%%%%%%
%\section{Results}

\section{Background}
\label{sec:background}

In this section, we provide all the theoretical dynamical decoupling background necessary in order to understand the experimental results presented in \cref{sec:expt-design}. In particular, after defining our Hamiltonian error model, we explain how different flavors of encoded DD, i.e., DD implemented using the stabilizers and/or logical operators of a QEC code, suppress the different types of errors that arise in our error model.

\subsection{Error model}

Let $\Ugrp(m)$ denote the group of unitary operators on $\mathbb{C}^m$. Let $\mP_n = \{\pm P_j, \pm i P_j\}_{j = 0}^{4^n-1} \subset \Ugrp(2^n)$ be the full Pauli group with $4^{n+1}$ elements
generated by $\{I,X,Y,Z\}^{\otimes n}$ and the phase factors 
$\{\pm 1,\pm i\}$, where $X\equiv \sigma^x$ and likewise for $Y$ and $Z$. There is a natural projection (quotient homomorphism) $\pi:\mathcal{P}_n \to \tilde{\mathcal{P}}_n$ which takes the Pauli operator and ignores its global phase, yielding a ``phase-stripped'' Pauli. Correspondingly, let $\tilde{\mP}_n = \mP_n / \{\pm 1,\pm i\} \subset \Ugrp(2^n) / \Ugrp(1)$
be the Abelian factor group with $4^n$ elements,
also known as the projective Pauli group. 

Consider the most general ``total decoherence'' system-bath interaction for $n$ qubits:
\beq
\label{eq:H_SB-Pauli}
H_{SB}=\sum_{{j=1}}^{4^n-1} \alpha_j  P_{j} \otimes B_j .
\eeq
In writing this expression we exclude the bath Hamiltonian $I_S\otimes B_0$ (where $B_0 = H_B$) and select exactly one representative $P_j$ from each of the $4^n$ equivalence classes $\tilde{\mathcal{P}}_n$ such that (s.t.) $P_0 = I_S$ (we use the notation $I_S$ to denote the identity operator on the entire system Hilbert space; we use $I$ for a generic identity operator). There is still some freedom in how we order the elements of $\tilde{\mathcal{P}}_n$ and which phases we choose. We use this freedom later to organize $\{P_j\}_{j=1}^{4^n-1}$ in the context of $[[n, k, d]]$ codes.

Our model for $H_{SB}$ is general enough to contain system-only error terms, since any of the bath operators $B_j$ may be set equal to the bath identity operator $I_B$. Henceforth, when unambiguous, we use $I$ to denote the identity operator regardless of the space on which it acts or its dimension. The joint system-bath free-evolution unitary is $f_\tau = e^{-i\tau (H_{SB} + I\otimes H_B)}$. The adjective ``free'' refers to the fact that we are leaving out any system-only control Hamiltonian; such control will enter later through DD or QEC. The goal of the latter is, respectively, to suppress or correct the deleterious effects caused by $H_{SB}$.

\subsection{Dynamical decoupling and quantum error correction}

In its simplest form, accounting only for first-order decoupling and ignoring pulse errors, DD theory can be understood as group symmetrization~\cite{Zanardi1999fk,Zanardi:99d,Viola:2000:3520}. Consider a discrete group $\mG \subset \Ugrp(2^n)$ with elements $\{g_j\}_{j=0}^{|\mG|-1}$ representing unitary transformations $g_j$ acting purely on the system. We set $g_0 \equiv g_{|\mG|}\equiv I_S$ and refer to the corresponding phase-stripped group $\tmG = \pi(\mG) = \left\{\tg_j\right\}_{j=0}^{K-1}\subset \Ugrp(2^n) / \Ugrp(1)$ as the ``decoupling group'', where $K\equiv |\tmG| = |\mG|/4$.
This yields the DD cycle
\beq
\label{eq:cycle}
\begin{aligned}
U(T)&=\prod_{j=0}^{K-1} \tg_j^{\dagger} f_\tau \tg_j = e^{-i T (\langle H_{SB}\rangle_{\mG}+I\otimes H_B)} + O(T^2) \\
&= p_K f_\tau p_{K-1} f_\tau \cdots f_\tau p_1 f_\tau .
\end{aligned}
\eeq
Here, $\{p_j = \tg_j\tg_{j-1}^\dagger\}_{j=1}^{K}$ are the pulses of the corresponding DD pulse sequence, $\tau$ is the pulse interval, and $T=K \tau$ is the total duration of one repetition of the sequence. Note that, equivalently, $\tg_{j-1} = p_{j-1}\cdots p_0$, where $p_0\equiv I_S$.

$\langle H_{SB}\rangle_{\mG}$ is the group-symmetrized system-bath Hamiltonian, where 
\beq
\label{eq:G-ave}
\langle A\rangle_{\mG}
\equiv
\frac{1}{|\mG |} \sum_{g\in\mG } g^{\dagger} A g = 
\frac{1}{K} \sum_{\tg\in\tmG} \tg^{\dagger} A \tg
\eeq
is the projection of the arbitrary bounded system-bath operator $A$ onto the subalgebra of operators that commute with every element of $\mG$ (i.e., its commutant). {$\langle A\rangle_{\mG}$ is also called the group average of $A$ with respect to $\mG$.}
\begin{mydefinition}[Decoupling]
\label{def:decoupling}
Let $A$ be an arbitrary operator acting on the joint system-bath Hilbert space. $\tmG$ decouples $A$ (to first order) if $\langle A\rangle_{\mG} = cI\otimes B$ where $c$ is a constant, including zero, and $B$ is an arbitrary bath operator. 
\end{mydefinition}

For example, $\tmG = \tilde{\mP}_n$ decouples an arbitrary $n$-qubit system-bath Hamiltonian since then $\langle H_{SB}\rangle_{\mG}=0$, albeit at a sequence time cost of $T=4^n\tau$~\cite{Viola1999PRL}. To illustrate this, consider $n=1$; the most general system-bath Hamiltonian of a single qubit is $H_{SB}=\sum_{{i=1}}^{3}\alpha_i P_i\otimes B_i$ with $P_i \in\mP_1\setminus I$.
If $\tmG={\tilde{\mP}_1}=\<X,Z\> =\{I,X,Y,Z\}$, then $U(T)$ simplifies into the well-known (universal) XY4 sequence $U(4\tau) = Y f_\tau X f_\tau Y f_\tau X f_\tau$~\cite{Maudsley1986ty}, and indeed, $\langle H_{SB}\rangle_{\mG}=0$.  The XX sequence is $U(2\tau) = X f_\tau X f_\tau$, and decouples $Y$ and $Z$, but not $X$ errors. The XY4 sequence can be seen as a concatenation of the XX and ZZ sequences, with the latter defined similarly to XX~\cite{Khodjasteh:2005xu}.  

Note that a common misconception is that DD is not effective against purely Markovian noise; however, this is not the case, essentially since even in the Markovian limit, the bath can have a nonzero correlation time~\cite{Szczygielski:2015aa,Addis:2015aa,Arenz:2018aa,Mozgunov:2019aa}.

In general, decoupling using subgroups of $\tilde{\mP}_n$ will eliminate parts of $H_{SB}$, presenting an opportunity to selectively combine DD with QEC~\cite{KhodjastehLidar:03,Ng2011PRA}. See \cref{app:QECC} for pertinent background on QEC codes. 

Let $\mathcal{S}$ be a stabilizer group specifying an $[[n, k, d]]$ stabilizer code. The group of logical operators is $\NS / \mathcal{S}$, where $\NS$ is the normalizer of $\mS$ in $\mP_n$ (which here coincides with the centralizer, i.e., the group of Pauli operators commuting with all elements of $\mS$).
We can pick a canonical set of generators of the stabilizer group and a canonical group of logical operators $\mathcal{L}$.\footnote{That is, any transversal for $\mS$ in $\NS$ satisfying $\pi^{-1}(\pi(\mL))=\mL$ can be picked as $\mL$, i.e., any subgroup of $\NS$ with exactly one representative from each coset $L\mS \subset \NS$ s.t. $iI \in \mL$.}
Any such $\mathcal{L}$ is isomorphic to $\mathcal{P}_k$ and can be thought of as a group of Paulis acting on logical qubits.

We can use the remaining freedom mentioned above to reorder $\{P_j\}_{j=0}^{4^n-1}$ and pick the phases of those operators to ensure that
\begin{itemize}
    \item $P_0 = I_S$
    \item $\mathcal{S}_{\star} = \{P_1, \dots, P_{2^{n-k}-1}\} = \mathcal{S} \setminus \{I_S\}$ (these are non-identity stabilizers acting trivially on the code space)
    \item $\mathcal{L}_{\star} = \{P_{2^{n-k}}, \dots, P_{2^{n+k}-1}\} \subset \mathcal{N}(\mathcal{S}) \setminus \pi^{-1}(\tilde{\mathcal{S}})$ (these manifest as undetectable logical errors when acting on the code space)
    \item $\mathcal{D}_{\star} = \{P_{2^{n+k}}, \dots, P_{4^{n}-1}\} \subset \mathcal{P}_n \setminus \mathcal{N}(\mathcal{S})$ (these are detectable errors).
\end{itemize}
We chose the phases of the representatives of $\tilde{\mathcal{S}}$ to ensure they are actually in $\mathcal{S}$ (and not different from elements of $\mathcal{S}$ by a phase).
This represents $\{P_j\}_{j=0}^{4^n-1}$ as a disjoint union
\begin{equation}
  \{P_j\}_{j=0}^{4^n-1} = \{I_S\} \dot\cup \mathcal{S}_{\star} \dot\cup \mathcal{L}_{\star} \dot\cup \mathcal{D}_{\star}.
\end{equation}

The system Hilbert space $\mH$ can be split as 
\beq
\mH  = \mH _{\text{stab}}\otimes\mH_{\text{log}},
\label{eq:H-decomp}
\eeq
where $\mathcal{S}$ acts on the first component and $\mathcal{L}$ acts on the second, and 
 where $\dim(\mH _{\text{stab}})=2^{n-k}$ and $\dim(\mH _{\text{log}})=2^{k}$, corresponding to the $n-k$ (virtual) syndrome qubits and $k$ logical qubits, respectively. 

With respect to this virtual tensor product decomposition induced by the syndrome measurement~\cite{ZLL:04},
every Pauli operator $E \in \mS_\star$, $\mL_\star$, or $\mD_\star$ can be written as:
\begin{itemize}
\item $E= P\otimes I\in \mS_\star$ with $P\in\mP_{n-k}$ (a stabilizer with trivial action on the code space)
\item $E= P\otimes Q\in \mL_\star$ with $[P,\mS_\star]=0$ (a logical operator manifesting as an error if it appears in $H_{SB}$).
\item $E = P\otimes Q\in \mD_\star$ with $\{P,S_i\}=0$ for one or more $S_i\in\mS_\star$ (a detectable error).
\end{itemize}
Correspondingly, it is always possible to write the system-bath interaction [\cref{eq:H_SB-Pauli}] as 
\bes
\label{eq:H_SB}
\begin{align}
H_{SB} &= \HSBstab + \HSBlog + \HSBdet \\
&\HSBstab \!\!= \sum_{i=1}^{2^{n-k}-1}  S_i\otimes B^{\text{stab}}_i\ , \quad S_i\in\mS_\star\\
\label{eq:H_SB-log}
&\HSBlog = \sum_{i=1}^{2^{n-k}(4^{k}-1)}  L_i\otimes B^{\text{log}}_i\ , \quad L_i\in\mL_\star\\
&\HSBdet = \sum_{i=1}^{4^n-2^{n+k}}  D_i\otimes B^{\text{det}}_i\ , \quad D_i\in\mD_\star ,
\end{align}
\ees
where $\{B^{\text{stab}}_i,B^{\text{log}}_i,B^{\text{det}}_i\}$ are bath operators with units of energy.

\cref{eq:H_SB} groups terms with similar action on the code space $\mC$. In total, $H_{SB} + I_S\otimes H_B$ has $4^n$ linearly independent terms corresponding to the elements of $\tilde{\mP}_n$.
Specifically, $\HSBstab+ I_S\otimes H_B$ collects all $2^{n-k}$ distinct terms in $\{I_S\} \dot\cup \mathcal{S}_{\star}$ that act as the identity operator on the code space and, therefore, are harmless. 
$\HSBlog$ collects all $2^{n-k}(4^{k}-1)$ Pauli operators in $\mL_\star$; these terms are non-trivial logical operators that leave the code space (and each of the other syndrome spaces) invariant. 
Finally, $\HSBdet$ collects all remaining $4^n-2^{n-k}-2^{n-k}(4^{k}-1)=4^n-2^{n+k}$ terms in $\mD_\star$, which correspond to 
detectable {and potentially correctable) errors.

We can now establish when an error $E \in \mP_n$ is decoupled. 
\begin{mylemma}
\label{lem:1}
Let $\mH  = (\mathbb{C}^2)^{\otimes n}$ be the Hilbert space of $n$ qubits. Then 
%$\mC  \subset \mH $ and 
all elements of $\mP_n$ 
%(and, hence, of $\mL  \subset \mP_n$ and $\mS  \subset \mP_n$) 
are unitary operators on $\mH$. 
%We let $\mG  = \mL  = \{\pm g_j, \pm i g_j\}_{j=0}^{4^k-1}$. 
Consider a term in $H_{SB}$ corresponding to some $E \in \mP_n$. It is decoupled by $\tmG$ in the sense of \cref{def:decoupling} if and only if there exists $g \in \mG $ anticommuting with $E$.
\end{mylemma}

\begin{proof}
If there exists $g \in \mG $ that anticommutes with $E$ then $g^{\dagger} E g = -E$ and
\begin{equation}
\langle E \rangle_{\mG}
=
\frac{1}{\abs{\mG}} \sum_{h\in\mG } h^{\dagger} E h = -\frac{1}{\abs{\mG}} \sum_{h\in\mG } (gh)^{\dagger} E {gh} = -\langle E \rangle_{\mG},
\end{equation}
so the group average $\langle E \rangle_{\mG} = 0$. If there is no such $g$, then all elements of $\mG $ commute with $E$ (because every two elements of $\mP_n$ either commute or anticommute). Then $g^{\dagger} E g = E$ for all $g \in \mG $, hence $\langle E \rangle_{\mG} = E$ (i.e., the term is undecoupled). 
\end{proof}

\subsection{Encoded dynamical decoupling}

A key observation is that the decoupling group is arbitrary~\cite{Viola:2000:3520,Zanardi:99d} and may, in particular, be formed from the stabilizer group $\mS$ and/or the group of logical operators $\mL$ of a stabilizer code $\mC$~\cite{PazSilva2013SciRep}. We call these different choices ``encoded dynamical decoupling,'' as they all involve the use of encoded operations (relative to a QEC code) as DD pulses, instead of physical DD pulses. We next provide several examples of encoded DD groups.

When using a QEC code, the approach of using the full Pauli group $\mP_n$ is excessive, since there is no need to decouple the terms in $\HSBstab$; this observation motivated Ref.~\cite{PazSilva2013SciRep} to introduce SLDD, which achieves a reduction from a DD group order of $4^n$ to $2^{n+k}$. The savings are substantial when $k\ll n$.

\begin{mydefinition}[SLDD]
For any stabilizer group $\mS$ and associated canonical group of logical operators $\mL$, the stabilizer-logical dynamical decoupling (SLDD) group is
\beq
\tmG_{\text{SLDD}} = \pi(\mS\mL) .
  \label{eq:SLDD}
\eeq
\end{mydefinition}
Note that $\mS\mL = \NS$ and thus independent of the choice of $\mL$.

The main weakness of SLDD is that it still incurs a cost that is exponential in the number of physical qubits $n$. This observation motivates us to consider a different hybrid approach that shifts more of the burden for handling correctable errors to the QEC code and uses only the logical operators of the code as DD pulses:

\begin{mydefinition}[LDD]
Pick a canonical group of logical operators $\mL$.
The logical dynamical decoupling (LDD) group is
\beq
\tmG_{\text{LDD}} \equiv \tmL = \pi(\mL) .
  \label{eq:LDD}
\eeq
\end{mydefinition}
Compared to SLDD, this approach has the desired effect of reducing the DD group order to $T\le 4^{k}$, independent of $n$.

For the sake of completeness, we also mention the case of DD using only the stabilizers:
\begin{mydefinition}[SDD]
For any stabilizer group $\mS$, the SDD group is
\beq
\tmG_{\text{SDD}} \equiv \tmS = \pi(\mS) .
  \label{eq:SDD}
\eeq

\end{mydefinition}
SDD can be seen as an alternative to QEC since it decouples all detectable errors. However, it does not decouple any logical errors and is therefore not a particularly useful method 
when used in conjunction with 
error correction.
It does play an important role in scenarios where QEC is not viable, such as in adiabatic quantum computing~\cite{PhysRevLett.100.160506,PhysRevA.86.042333}.

We can now classify the different decoupling groups by the errors they decouple.

\begin{mytheorem}[Decoupling by group type]
\label{lem:2}\quad
\begin{enumerate}
\item
SDD decouples all detectable errors: $\ave{\HSBdet}_{\tmG_\text{SDD}} = 0$.
\item 
LDD decouples all logical errors: $\ave{\HSBlog}_{\tmG_\text{LDD}} = 0$.
\item 
SLDD decouples all errors: $\ave{\HSBlog+\HSBdet}_{\tmG_\text{SLDD}} = 0$.
\end{enumerate}
\end{mytheorem}

\begin{proof}
It follows from \cref{lem:1} that it suffices to show that for each term in $\HSBstab$, $\HSBlog$, or $\HSBdet$, there exists $g \in \mG $ it anticommutes with, where $\mG$ is the corresponding decoupling group. Indeed:
\begin{enumerate}
\item SDD: Each detectable error anticommmutes with at least one element of $\mS$.
\item LDD: Each nontrivial logical error anticommmutes with at least one element of $\mL$.
\item Since $I\in \mL$, the stabilizer $\mS \subset \mG_{\text{SLDD}} = \bigl\{\mS L \mid L\in\mL \bigr\}$. Thus, the group average of $\HSBlog+\HSBdet$ with respect to $\mG_{\text{SLDD}}$ includes both $\mS$, which decouples all detectable errors, and $\mL$, which decouples all logical errors.
\end{enumerate}
\end{proof}

\cref{tab:comp} summarizes the relative resource cost of full Pauli group DD, {SDD}, LDD, and SLDD.}

\begin{table}[h]
\begin{tabular}{|p{2cm}||p{2cm}|p{2cm}|p{2cm}|}
 \hline
 \multicolumn{4}{|c|}{Comparative DD resource cost} \\
 \hline
 Decoupling group $\mG$& \# of pulses & error type suppressed & requires QEC\\
 \hline
 $\tilde{\mP}_n$   & $2^{2n}$    &all &  No\\
SDD&   $2^{n-k}$  & detectable & No\\
LDD & $2^{2k}$ & logical & Yes\\
SLDD&   $2^{n+k}$  & detectable \& logical & No\\
  \hline
\end{tabular}
\caption{The comparative resource cost for various DD methods in terms of the number of pulses required to suppress the most general system-bath Hamiltonian $H_{SB}$ [\cref{eq:H_SB}] acting on $n$ qubits, and whether active error correction is required in addition.  Full Pauli group DD and SLDD suppress all error mechanisms that can decohere code states, while LDD does not, as it leaves dealing with these error mechanisms for the QEC cycle. Consequently, LDD does not scale with $n$ but with $k$, whereas Pauli group DD and SLDD both scale with $n$.}
\label{tab:comp}
\end{table}

It is common to use only subgroups of the complete group of logical operators. In the context of standard (unencoded) DD, this is the case, e.g., when using single-axis sequences such as XX instead of the universal XY4 sequence. In the present context, we would like to consider subgroups of $\NS$ other than $\mS$ and $\mL$, which will similarly suppress selected subsets of errors. We refer to this as normalizer DD, since the corresponding DD group elements are generic normalizer elements, as opposed to canonical logical operators.

\begin{mydefinition}[NDD]
 Normalizer DD (NDD) is any pulse sequence formed from using a subgroup of $\tmG_{\text{SLDD}}$ other than $\tmG_{\text{SDD}}$ or $\tmG_{\text{LDD}}$.
\end{mydefinition}
NDD arises in our experiments, as described in \cref{sec:expt-design}. We also discuss an example in the next subsection.

It is important to emphasize that additional errors may and will generally be decoupled by each of the decoupling groups, beyond those mentioned in \cref{lem:2}. For example, a detectable weight-$1$ Pauli operator that anticommutes with an element of $\mG_\text{LDD}$ will be decoupled despite not being a logical error for any code with $d\ge 2$. 

Another important point is that since we are employing a Hamiltonian noise model, any terms that arise in second order perturbation theory, i.e., the $O(T^2)$ terms we are neglecting in \cref{eq:cycle}, will mix $\HSBstab$, $\HSBlog$, and $\HSBdet$ through commutators. This means that the clean separation we have assumed between error types disappears at this higher order of perturbation theory. However, higher-order decoupling sequences are known that achieve suppression up to arbitrary order $q$ [i.e., leaving an $O(T^{q+1})$ unsuppressed error term in $U(T)$]~\cite{Khodjasteh:2005xu,Uhrig:2007qf,West:2010:130501,Wang:10,Xia:2011uq}. For simplicity, in this work we restrict our attention to first-order sequences.

\subsection{Illustration of LDD and NDD: the \texorpdfstring{$[[4,2,2]]$}{[[4,2,2]]} code}
\label{sec:illustr}

To illustrate the concepts discussed above, we analyze the example of the $[[4,2,2]]$ code, as it plays a key role in our experiments. 

The stabilizer group of the $[[4,2,2]]$ code is $\mS  = \langle XXXX, ZZZZ \rangle$. 
A canonical set of logical operators for the code can be chosen such that $\overline{X}_1=XIIX$, $\overline{X}_2=IIXX$, $\overline{Z}_1=IIZZ$ and $\overline{Z}_2=ZIIZ$,
i.e., $\mL  = \langle i, XIIX, IIXX, IIZZ, ZIIZ \rangle$ (see \cref{app:422} for additional pertinent details regarding this code). Thus, $\{XIIX, IIXX, IIZZ, ZIIZ\}$ is the generator set of the $16$-element LDD group $\tmG_{\text{LDD}} = \{\tg_j\}$. This LDD group (as well as any of the corresponding NDD groups) 
suppresses every logical error $E_{\text{log}}$ in the $2^{n-k}(4^{k}-1) = 4\times 15=60$-element group $(\NS\setminus\mS)  / \{\pm 1, \pm i\}$: the group-averaged logical system-bath Hamiltonian $\langle \HSBlog\rangle_{\mG}$ [\cref{eq:H_SB-log}] vanishes, as every one of its terms is a logical operator that anticommutes with an element of $\mG$.

This leaves $4^n-2^{n+k} = 256-64 = 192$ detectable errors. These errors can be written compactly as $\mD = \{(XIII\cdot\tNS)\cup (YIII\cdot\tNS) \cup (ZIII\cdot\tNS)\}$, yielding $3\times |\tNS| = 192$ terms.
Each such error is detectable because it anticommutes with at least one element of $\mS$. Of these, only the $12$ errors of the form $PQPP$, with $P\ne Q\in\{I,X,Y,Z\}$, are unsuppressed by LDD. 

Now suppose that instead of suppressing the full $\HSBlog$, the goal is to suppress just $H_{SB}^{\overline{Z}}\equiv \overline{Z}_1\ox B_1^{\text{log}} + \overline{Z}_2\ox B_2^{\text{log}}$. This scenario is particularly relevant to the experiments we describe below. In this case, it suffices to use the smaller LDD-subgroup $\langle \overline{X}_1\overline{X}_2\rangle$, since both $\overline{Z}_1$ and $\overline{Z}_2$ anticommute with $\overline{XX}\equiv\overline{X}_1\overline{X}_2$. The corresponding LDD cycle is then $\overline{XX}f_\tau\overline{XX}f_\tau$, where $f_\tau = \exp(-i\tau H_{SB}^{\overline{Z}})$. For crosstalk-related reasons that will become apparent below, it turns out that it is advantageous to replace the second $\overline{XX}$ by $S_1\overline{XX}  = IXIX \equiv \overline{X'X'}$, where $S_1=XXXX\in\mS$. This replacement yields the $4$-pulse NDD cycle $\overline{X'X'}f_\tau\overline{XX}f_\tau \overline{X'X'}f_\tau\overline{XX}f_\tau = (\overline{X'X'} f_\tau \overline{X'X'}) (S_1 f_\tau S_1) (\overline{XX}f_\tau \overline{XX}) (If_\tau I)$. This cycle is generated by the NDD group $\tmG = \<\overline{XX},S_1\> = \{I,\overline{XX},S_1,\overline{X'X'}\}$. Below, we refer to this decoupling group as NXX.

As another example of NDD, consider the encoded XY4 sequence with the LDD group $\tmG = \<\overline{XX},\overline{ZZ}\>$ and DD cycle $\overline{YY}f_\tau\overline{XX}f_\tau\overline{YY}f_\tau\overline{XX}f_\tau$. Again, for crosstalk-related reasons it will turn out to be advantageous to replace the canonical $\overline{YY}=YIYI$ by $\overline{Y'Y'}=S_2\overline{YY}$, where $S_2=YYYY\in\mS$. The resulting DD cycle $\overline{Y'Y'}f_\tau\overline{XX}f_\tau\overline{Y'Y'}f_\tau\overline{XX}f_\tau$ is generated by the NDD group $\tmG = \<\overline{XX},\overline{Z'Z'}\> = \{I,\overline{XX},\overline{Y'Y'},\overline{Z'Z'}\}$, where $\overline{Z'Z'}=S_2\overline{ZZ}$. Below, we refer to this decoupling group as NXY4.

%%%%%%%%%%%%%%%%%%%%%%%%%%%%%%%%%%%%
\begin{figure*}[ht]
\hspace{0cm}{\includegraphics[scale=.66]{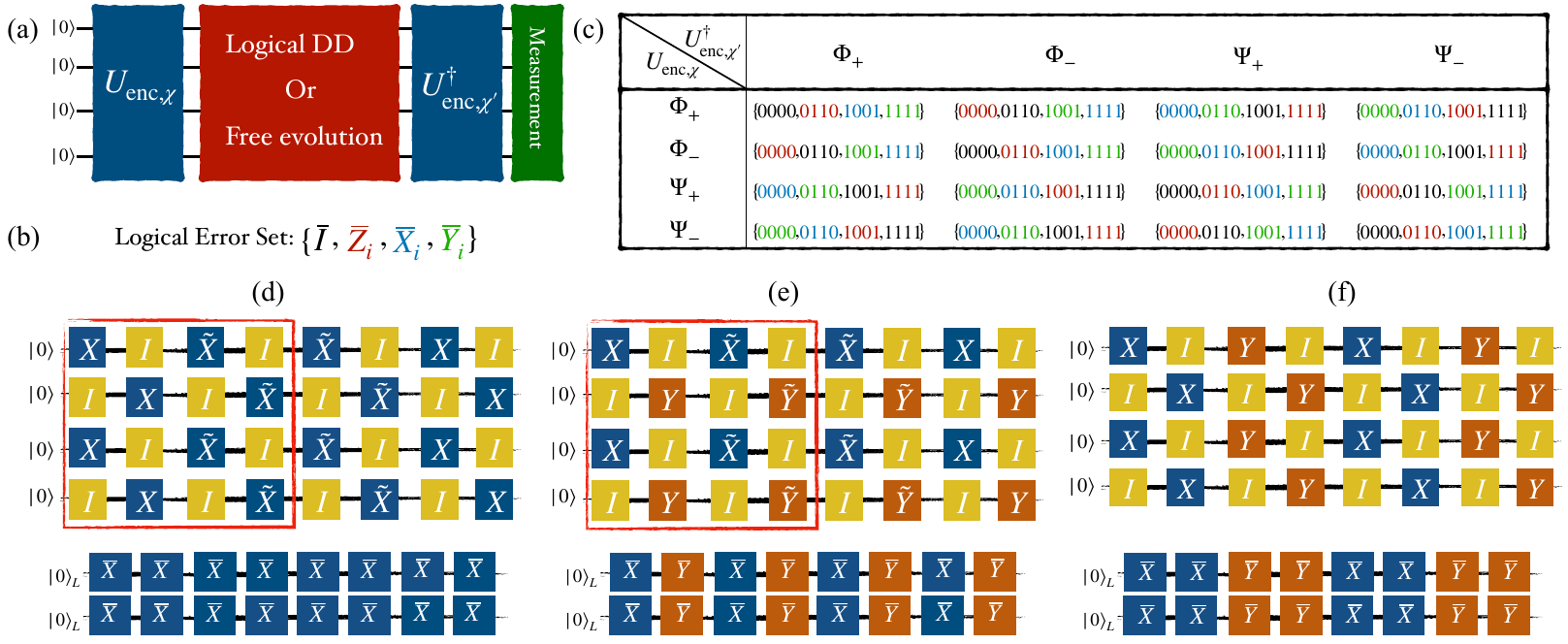}} 
\caption{(a) Schematic of the experimental design: we encode a logical Bell state $\chi$ using the $[[4,2,2]]$ code and let it undergo either free evolution or NDD for a duration $\tau$. We then run the unencoding circuit for $\chi'$ and measure the resulting bitstring. The encoding and unencoding circuits together take up to $5\mu$s. (b) The logical error set that relates the four Bell states as a result of a logical operator acting on one ($i=1$ or $2$) of the two logical qubits. (c) Interpretation of different bitstrings for an encoded state ($U_{\rm{enc},\chi}$), depending on the unencoding ($U_{\rm{enc},\chi^\prime}^\dagger$) circuit used. The color code indicates that the encoded state has undergone either no error (black) or one of the logical errors (red, blue, green) indicated in (b). The DD sequences used in our experiments are (d) {Robust Normalizer XX (RNXX), (e) Robust Normalizer XY4 (RNXY4)}, and (f) physically Staggered XY4 (SXY4). Each panel shows both the physical implementation (top) and the logical interpretation of each sequence (bottom). A dagger in (d) and (e) denotes a negative $\pi$ rotation, i.e., $X = R_x(\pi) = \exp(-i\pi \sigma^x/2)$, $\tilde{X} = R_x(-\pi) = \exp(i\pi \sigma^x/2)$, and likewise for $Y$. {In (d) the first column is the canonical $\overline{XX} = XIXI$ (see \cref{app:422}), the second column is $\overline{X'X'}=S_1\overline{XX}$, where $S_1=XXXX\in\mS$, etc.
%, so the full logical sequence is an sNDD sequence as depicted in the bottom panel. 
Likewise, in (e), the second column is $\overline{Y'Y'}=S_2\overline{YY}$, where the canonical $\overline{YY}=YIYI$ and $S_2=YYYY\in\mS$}.} 
\label{fig:schematic}
\end{figure*}
%%%%%%%%%%%%%%%%%%%%%%%%%%%%%%%%%%%%

%%%%%%%%%%%%%%%%%%%%%%%%%%%%
%%%%%%%%%%%%%%%%%%%%%%%%%%%%
%%%%%%%%%%%%%%%%%%%%%%%%%%%%
\section{Experimental design}
\label{sec:expt-design}

For our experiments, while sequences utilizing the full LDD/NDD group can be constructed (see \cref{app:full-LDD}), we instead use a subgroup-NDD, generated by a subgroup of $\mL$ and a subgroup of $\mS$. This subgroup NDD works better in practice because of its robustness and ability to suppress crosstalk. This section is dedicated to demonstrating experimentally how our hybrid QED + subgroup-NDD protocol results in a significant fidelity enhancement of entangled logical qubits.

The data for the experiments we report here were collected on two separate occasions from a total of $24$ sets of four-qubit experiments run on the \texttt{ibm\_kyiv}~\cite{IBMQ} quantum processor. Dataset 1, using $14$ sets of qubits, was collected during the week of August 12$^{\rm{th}}$ (2024). Dataset 2, using $10$ sets of qubits, was collected on March 1\textsuperscript{st} and 2\textsuperscript{nd} (2025). These $10$ four-qubit sets differ from those in dataset 1 in order to test the robustness of our results. The figure captions below specify the dataset. For each data point in our results, fidelity is independently calculated for each qubit set and then bootstrapped by resampling. The mean fidelity and standard deviation are derived from the bootstrapped data, where larger error bars indicate greater variability among the qubit sets.

The \texttt{ibm\_kyiv} processor consists of coupled, fixed-frequency transmons~\cite{transmon-invention}. Such qubits exhibit an always-on interaction between adjacent pairs~\cite{BlaisRMP}. This crosstalk gives rise to weight-2 error terms that correspond to logical errors with our $[[4,2,2]]$ code choice. Other errors of weight $\ge 2$, if present, would likewise correspond to logical errors. To demonstrate that NDD can suppress all logical errors, we design an experiment that enables the unequivocal detection of such errors.

To this end, we use the two logical qubits of the $[[4,2,2]]$ code to create logical Bell states $\ket{\Phi_\pm}=(\ket{\overline{00}}\pm\ket{\overline{11}})/\sqrt{2}$ and $\ket{\Psi_\pm}=(\ket{\overline{01}}\pm\ket{\overline{10}})/\sqrt{2}$. The corresponding encoding circuits $U_{\text{enc},\chi}$, where $\chi\in\{\ket{\Phi_\pm},\ket{\Psi_\pm}\}$, create two copies of physical Bell states, followed by the application of a logical controlled-NOT ($\overline{\text{CNOT}}$) to generate the logical Bell states (see \cref{app:422}). In principle, it would be possible to detect specific logical errors using other logical states (e.g., $\ket{\overline{00}}$ would allow us to detect $X$- or $Y$-type logical errors), but we specifically opt for logical Bell states as they have shorter encoding circuits. Additionally, the four logical Bell states form an orthonormal basis and the logical errors we are interested in permute this basis.

To estimate the fidelity of a logical Bell state $\chi$, we start from the physical ground state $\ket{0000}$, encode into the state of interest by applying $U_{\text{enc},\chi}$, and then let it evolve for some time $\tau$, either freely or subject to DD.
After time $\tau$ we unencode by applying $U_{\text{enc},\chi}^\dagger$, and measure all qubits.  Measuring the bitstring $0000$ would indicate that no error occurred.

If, instead, we obtain any of the odd-Hamming weight bitstrings, this would signal detection of a physical error, as such states are not in the codespace. The final possibility is that we obtain one of the other even-Hamming weight bitstrings $\{0110,1001,1111\}$, corresponding to a logical error. This procedure can be used to estimate the probability of finding the ground state as the empirical fraction of ground state measurement outcomes, which is also the fidelity of the logical Bell state $\chi$.

%%%%%%%%%%%%%%%%%%%%%%%%%%%%%%%%%%%%%%%%%%%%%%%%%%%%%%%%
\begin{table*}[ht]
\centering
\begin{tabular}{@{}lccccccc@{}}
\toprule
 & No DD & XY4 & UR$_{6}$ & UR$_{8}$ & UR$_{10}$ & UR$_{18}$ & RGA8${_a}$ \\ 
\midrule
$\ket{\Phi_+}$   & $65.44 \pm 0.24$ & $83.94 \pm 0.18$ & $83.08 \pm 0.19$ & $83.11 \pm 0.19$ & $82.05 \pm 0.19$ & $81.43 \pm 0.20$ & $\boxed{84.20 \pm 0.18}$ \\
$\ket{\Phi_-}$   & $61.23 \pm 0.24$ & $\boxed{83.78 \pm 0.18}$ & $81.92 \pm 0.19$ & $82.63 \pm 0.19$ & $81.07 \pm 0.19$ & $80.81 \pm 0.18$ & $83.34 \pm 0.18$ \\
$\ket{\Psi_+}$   & $66.18 \pm 0.23$ & $\boxed{84.68 \pm 0.18}$ & $84.17 \pm 0.18$ & $83.37 \pm 0.19$ & $82.46 \pm 0.19$ & $81.21 \pm 0.19$ & $84.65 \pm 0.18$ \\
$\ket{\Psi_-}$   & $65.43 \pm 0.23$ & $\boxed{86.57 \pm 0.17}$ & $85.20 \pm 0.18$ & $85.53 \pm 0.18$ & $84.17 \pm 0.18$ & $83.44 \pm 0.17$ & $86.38 \pm 0.17$ \\
\bottomrule
\end{tabular}
\caption{Fidelity ($\%$) of the encoded logical Bell states without and with DD, for a variety of pulse sequences. Experiments were performed on fourteen different sets of qubits on \texttt{ibm\_kyiv}, with 4000 shots per set. The fidelities were computed independently for each set and then bootstrapped. The reported values represent the bootstrapped averages with $1\sigma$ uncertainty. The boxed values are the highest fidelities in each row.}
\label{tab:tab-1}
\end{table*}
%%%%%%%%%%%%%%%%%%%%%%%%%%%%%%%%%%%%%%%%%%%%%%%%%%%%%%%%

To obtain the fidelity of a logical error, we first observe that each of the logical Bell states is related to the other three logical Bell states through a specific logical error: $\chi^\prime=\overline{O}\chi$, for $\overline{O}\in\{\overline{X}_i, \overline{Y}_i, \overline{Z}_i\}$ and $i=1$ or $2$; i.e., a logical operator that acts on either of the two logical qubits. More generally, any other combinations of logical weight-$1$ errors also map the input Bell state to the other Bell states. That is, for $\bar{p},\bar{q}\in\{\overline{X}, \overline{Y},\overline{Z}\}$, we have  $\chi^\prime = (\bar{p} \otimes \bar{q}) \chi = I\otimes(\bar p \bar q) \chi = (\bar p \bar q)\otimes I \chi$, up to overall phases. We note that the following two-qubit logical errors leave the logical Bell states unchanged and therefore do not influence the experimental outcome:  $\overline{O}\chi=\chi$, for $\overline{O}\in\{II, \overline{X}\overline{X}, \overline{Y}\overline{Y}, \overline{Z}\overline{Z}\}$.

We thus proceed as follows: instead of applying $U_{\rm{enc},\chi}^\dagger$, at $t=\tau$ we deliberately unencode into a different logical Bell state $\chi^\prime$ [see \cref{fig:schematic}(a)]. In this manner, each of the even-weight bitstrings gives us a measure of the occurrence of one of the logical errors $\mathcal{E}_{\overline{O},\tau}$, where $\overline{O}$ is determined by the unencoding we choose [see \cref{fig:schematic}(b,c)]. Using this methodology, we can detect the occurrence of different types of logical errors and quantify the associated error probability.

In more detail, each row in \cref{fig:schematic}(c) shows a different initially encoded logical Bell state. Each column corresponds to one of the unencoding circuits. In each unencoding scenario, the different bitstrings indicate that the state being unencoded is either the initial logical Bell state or some other logical Bell state. For example, consider the case where we originally encode $\ket{\Phi_+}$ and then unencode into $\ket{\Phi_-}$ at $t=\tau$. If the state being unencoded is indeed $\ket{\Phi_-}=\overline{Z}_i\ket{\Phi_+}$, then the $0000$ bitstring signals that a $Z$ logical error ($\overline{Z}_i$) has occurred. However, if no error has occurred, then due to unencoding into $\ket{\Phi_-}$, the result should be the $0110$ bitstring. Generalizing, it is possible to detect the different logical errors using the experiments indicated in \cref{fig:schematic}(c). 

However, we specifically choose the unencoding so that it is always the $0000$ bitstring that corresponds to the occurrence of the logical error operator in which we are interested. We make this choice since $\ket{0000}$ is the ground state of the system and therefore is robust against relaxation errors. This strongly increases the likelihood that the detected errors are purely logical and are unaffected by thermal relaxation. Note that due to this choice, the bitstring that corresponds to the fidelity $\mathcal{F}_\chi(t)$ of the prepared initial state varies [black color-coded bitstrings in \cref{fig:schematic}(c)]. 

From here on, we use the notation $\chi\mapsto\chi'$ to denote the procedure of preparing the encoded logical Bell state $\chi$ and unencoding into $\chi'$, i.e., of using the encoding unitary $U_{\text{enc},\chi}$ and the unencoding unitary $U^\dagger_{\text{enc},\chi'}$.

We note that logical state tomography is an alternative method for learning about the performance of LDD or NDD, and in particular, for certifying the entanglement of logical Bell states. However, the $[[4, 2, 2]]$ code imposes some additional challenges in performing logical tomography, which precluded its use in our work; see \cref{app:tomo} for details.

%%%%%%%%%%%%%%%%%%%%%%%%%%%%%%%%%%%%
\begin{figure}[ht]
\hspace{0cm}{\includegraphics[scale=.42]{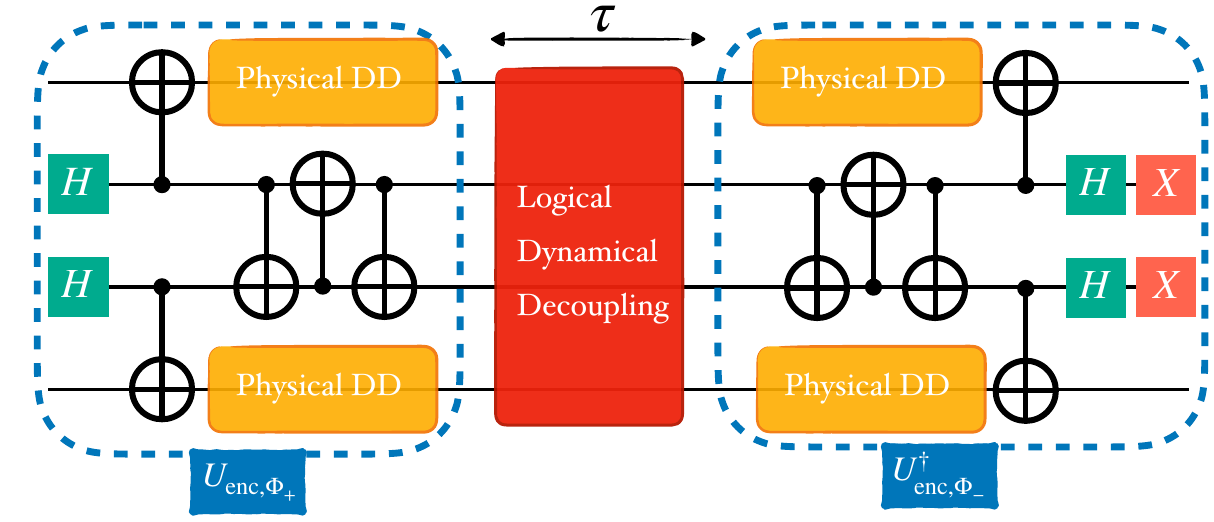}} 
\caption{Circuit schematic for encoding $\ket{\Phi_+}$ and unencoding $\ket{\Phi_-}$ with physical and logical DD.
%logical Bell states  with no delay time (see \cref{fig:schematic}(a)). 
Physical XY4 dynamical decoupling sequences (yellow boxes) are inserted into the idle gaps of the circuit. The code subspace is protected with NDD sequences. This is the DD-protected version of the circuit shown in \cref{fig:schematic}(a).}
\label{fig:circuit}
\end{figure}
%%%%%%%%%%%%%%%%%%%%%%%%%%%%%%%%%%%%

%%%%%%%%%%%%%%%%%%%%%%%%%%%%%%%%%%%%
\begin{figure*}[ht]
\hspace{0cm}{\includegraphics[scale=.66]{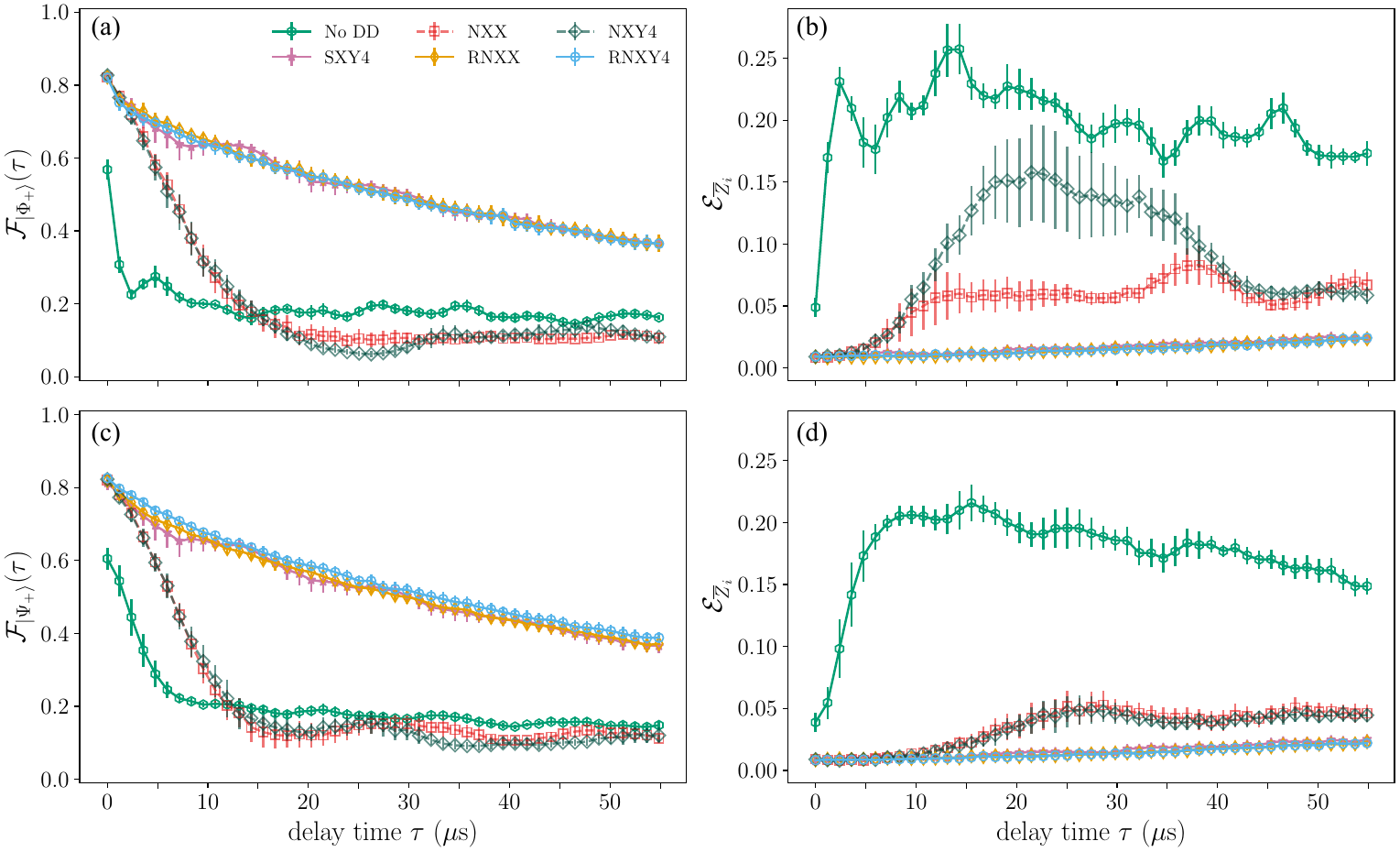}} 
\caption{Performance of NDD (dataset 1). The experiments encode the logical Bell states  $\ket{\Phi_+}$ in (a) and (b), and $\ket{\Psi_+}$ in (c) and (d). We then unencode in the logical Bell state related through $\overline{Z}_i$ on either of the logical qubits; i.e., $\ket{\Phi_+}\mapsto \ket{\Phi_-}$ in (a) and (b), and $\ket{\Psi_+}\mapsto \ket{\Psi_-}$ in (c) and (d). In this setting, the frequency of $0110$ outcomes [(a) and (c)] is a measure of the fidelity of the originally encoded state, while $0000$ corresponds to the detection of a $\overline{Z}_i$ error [(b) and (d)]. Without any DD, the fidelity is low [green in (a) and (c)] and the probability of logical errors is high [green in (b) and (d)]. 
Standard NDD sequences (NXX and NXY4; dashed light red and blue) improve the fidelity and logical error probability at short times ($\lesssim 15\mu$s) but their performance declines at longer times: they exhibit coherent errors as seen in the oscillations of the corresponding curves. Robust sequences (RNXX and RNXY4; solid dark red and blue) are the top performers; their ability to suppress the $\overline{Z}_i$ error is particularly noteworthy. Error bars are $1\sigma$ standard deviation after bootstrapping the data.} 
\label{fig:res-1}
\end{figure*}
%%%%%%%%%%%%%%%%%%%%%%%%%%%%%%%%%%%%

%%%%%%%%%%%%%%%%%%%%%%%%%%%%
\section{Experimental Results}
\label{sec:expt-results}

%%%%%%%%%%%%%%%%%%%%%%%%%%%%
\subsection{Physical dynamical decoupling improves logical Bell state fidelity}

We first show that we can substantially improve the logical Bell fidelity by padding the idle gaps of the encoding circuits with physical (as opposed to logical or normalizer) DD sequences. An \emph{idle gap} is a temporal circuit segment during which no gates are applied. Such gaps occur, e.g., when a pair of qubits is involved in a two-qubit gate that takes much longer than a single-qubit gate simultaneously being applied to another qubit; the latter is then idle after the completion of the single-qubit gate, while awaiting the completion of the two-qubit gate.

As shown in \cref{fig:circuit}, we inserted various DD sequences into the idle gaps of Bell state circuits. This includes XY4~\cite{Viola1999PRL}, universally robust sequences UR$_n$~\cite{Genov2017PRL} for $n=6,8,10, 18$, and RGA8$_a$~\cite{Quiroz2013PRA}.

Since the UR$_n$ sequences rely heavily on the use of both positive and negative rotations, it is important to clarify that henceforth, we use the notation $X \equiv R_x(\pi) = \exp(-i\pi \sigma^x/2)$, $\tilde{X}\equiv R_x(-\pi) = \exp(i\pi \sigma^x/2)$, and likewise for $Y$, to denote physical DD pulses. That is, unlike \cref{sec:background}, the symbols $X$ and $Y$ no longer denote the Pauli $\sigma^x$ and $\sigma^y$ operators. The reason this matters is that physical DD pulses suffer from control errors (both axis-angle and angle-magnitude errors), as well as errors due to the presence of the system-bath interaction while the pulse is on. In other words, the notation $R_x(\pi)$, etc., hides the fact that in reality the exact rotation axis deviates from $x$ and the exact rotation angle deviates from $\pi$. When these deviations are accounted for, DD sequences that are formally identical due to phase cancellation, such as $XX$ and $X\tilde{X}$, result in different operations at the physical level. For more details, see \cref{app:cross}.

We performed the experiments (during the week of August 12$^{\rm{th}}$ 2024) on fourteen different sets of qubits of \texttt{ibm\_kyiv}  and report the average fidelity values in \cref{tab:tab-1}. In these experiments, we encoded and unencoded the same state, i.e., used $U_{\rm{enc},\chi}$ along with $U_{\rm{enc},\chi}^\dagger$. We see that using DD, the encoding fidelity improves by $\sim20\%$ for all four logical Bell states, with XY4 and RGA8$_a$ being the top performers (we attribute the lower performance of the longer UR$_n$ sequences to pulse interference effects~\cite{Vezvaee2024arxiv}). 

In all, these results demonstrate that logical Bell state preservation can benefit significantly from physical DD. 
Having established the utility of physical DD, we proceed to combine it with NDD in the next section.

%%%%%%%%%%%%%%%%%%%%%%%%%%%%
%%%%%%%%%%%%%%%%%%%%%%%%%%%%
%%%%%%%%%%%%%%%%%%%%%%%%%%%%

%%%%%%%%%%%%%%%%%%%%%%%%%%%%%%%%%%%%
\begin{figure}[ht]
\hspace{0cm}{\includegraphics[scale=.55]{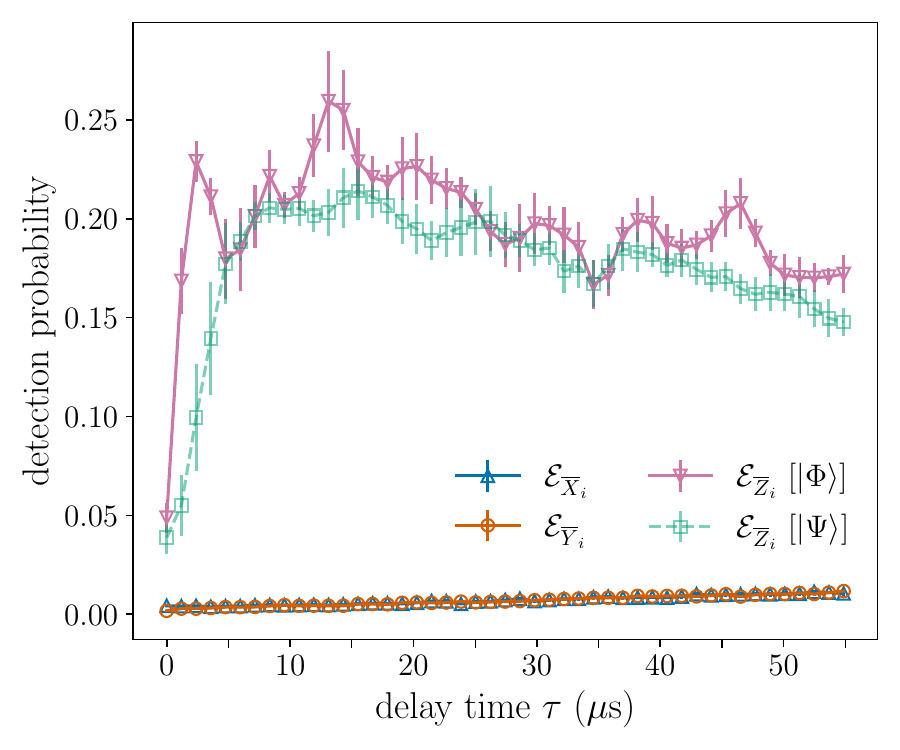}} 
\caption{Detection probability (i.e., probability of the $0000$ bitstring) of logical $X$, $Y$, and $Z$-type errors (dataset 1). The lower probability of logical $X$ and $Y$-type errors signals that the main source of logical errors is $ZZ$ crosstalk. These experiments are performed without DD.} 
\label{fig:res-2}
\end{figure}
%%%%%%%%%%%%%%%%%%%%%%%%%%%%%%%%%%%%

\subsection{Logical error suppression and detection by NDD}

For all our experimental results, ``No DD'' refers to $[[4,2,2]]$ encoding without any physical or logical DD.  In all other experiments, we use physical XY4 to pad all the encoding and unencoding idle gaps, and combine them with various flavors of NDD. This choice allows us to clearly assess the improvements introduced by DD, compared to relying solely on the code's error detection capabilities. 

\subsubsection{Logical $Z$ errors}
\label{sec:logZerr}

We start by gradually increasing the time delay $\tau$ (up to 55 $\mu$s) between the encoding and unencoding without DD. This situation is relevant in the context of QEC experiments. For example, one could prepare an encoded qubit and then leave it to idle while other logical operations are applied to other encoded qubits~\cite{Bluvstein2023Nature}. 

We perform the $\overline{Z}_i$ error detection using the logical Bell states $\ket{\Psi_{+}}$ and $\ket{\Phi_{+}}$ as discussed in \cref{sec:expt-design}, and display the results in \cref{fig:res-1}. As can be seen in \cref{fig:res-1}(a) and (c), without NDD the free evolution fidelity (denoted as No DD) decays rapidly and exhibits $ZZ$ crosstalk-induced oscillations. \Cref{fig:res-1}(b) and (d) show that logical $Z$ errors accumulate over time. 

We next use two types of NDD sequences to suppress this effect: Normalizer XX (NXX) and Normalizer XY4 (NXY4). Physical-level schematics of these NDD sequences are shown in the boxed four-pulse sequences of \cref{fig:schematic}(d) and (e), respectively (disregarding the tilde notation). Their logical-level counterparts are shown at the bottom of \cref{fig:schematic}. We generate both NXX and NXY4 using the native logical operations of the $[[4,2,2]]$ code. For example, in the upper part of \cref{fig:schematic}(d), reading the first {(second) column from top to bottom yields $XIXI = \overline{XX}$ ($\overline{X'X'}=IXIX = S_1\overline{XX}$; $S_1=XXXX$), which is the $\bar{X}\bar{X}$ ($\bar{X}'\bar{X}'$)} column at the bottom of \cref{fig:schematic}(d). Reading the pulse sequences from left to right, the staggering (appearance of delays as indicated by the identity operations) is deliberately introduced to suppress crosstalk at the physical level~\cite{Zhou2023PRL}; see \cref{app:cross} for more details.

The corresponding results are denoted as NXX (red) and NXY4 (dark green) in \cref{fig:res-1}. NXX is a single-axis non-universal 
%sNDD 
sequence, while NXY4 is ``nearly'' universal: it suppresses all logical errors except its own generating set $\{XIXI,IXIX,YIYI,IYIY\}$. At short times ($\lesssim 15\mu$s), these results are better than those without DD because some errors (including $ZZ$ crosstalk) are suppressed by both NXX and NXY4. However, at longer times, the benefit is lost and, moreover, small oscillations appear that indicate the presence of coherent errors~ \cite{tripathi2024DB}. To overcome this, we create robust versions of NXY4 and NXX by ensuring that all physical qubits in the code undergo physical DD sequences robust to pulse errors. Specifically, we use the universally robust (UR) sequence family~\cite{Genov2017PRL}, and ensure that each physical qubit undergoes a UR$_4$ sequence, i.e., $X\tilde{X}\tilde{X}X$ or $Y\tilde{Y}\tilde{Y}Y$, where a tilde denotes an $X$ or $Y$ rotation by $-\pi$ instead of $\pi$. These robust versions, which we call RNXX and RNXY4, are the full sequences shown in \cref{fig:schematic}(d) and (e). The performance of RNXX (yellow) and RNXY4 (blue), as seen in \cref{fig:res-1}, exhibits a significant improvement. Notably, the logical Bell state fidelities decay more slowly and without oscillations [panels (a) and (c)], and the logical $Z$ errors are strongly suppressed [panels (b) and (d)]. 

In addition, we apply the physical staggered XY4 (SXY4) sequence. This corresponds to applying a single XY4 sequence to each physical qubit but in a staggered manner to reduce crosstalk~\cite{Zhou2023PRL}. This sequence, shown in \cref{fig:schematic}(f), also corresponds to a non-universal purely logical-$Z$-error suppressing, NDD sequence for the $[[4,2,2]]$ code [bottom of panel (f)]. Interestingly, \cref{fig:res-1} shows that SXY4 performs on par with RNXX and RNXY4. This finding signals that the dominant logical errors are of $Z$-type. We confirm this in the following. 

%%%%%%%%%%%%%%%%%%%%%%%%%%%%%%%%%%%%
\begin{figure*}
\hspace{0cm}{\includegraphics[scale=.65]{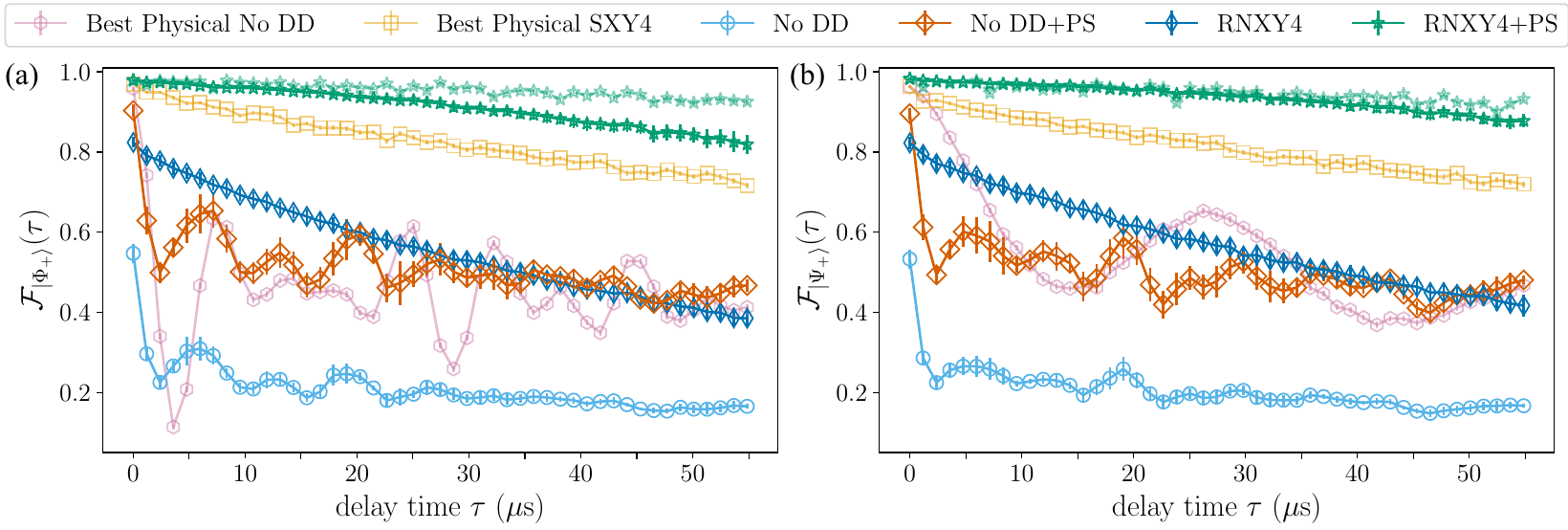}} 
\caption{Fidelity of logical Bell states (dataset 2) for (a) $\ket{\Phi_+}$ (using $\ket{\Phi_+}\mapsto \ket{\Phi_-}$) and (b) $\ket{\Psi_+}$ (using $\ket{\Psi_+}\mapsto \ket{\Psi_-}$). In both cases, without DD, the fidelity drops to 20$\%$ in $\approx 10\mu\rm{s}$ (No DD; light blue). Note that for \texttt{ibm\_kyiv},  median $T_1=279.92~\mu$s, median $T_2=111.926~\mu$s. Using error detection and applying postselection, the fidelity improves to $\approx 35\%$ (No DD+PS; brown). Using NDD alone is roughly equivalent to No DD+PS, but removes the crosstalk oscillations (RNXY4; dark blue). When we combine NDD with postselection, the fidelity increases substantially and remains above $82\%$ [in (a)] and $87\%$ [in (b)] over a 55$\mu$s period (RNXY4+PS; dark green). The light green dashed lines in each panel represent the results obtained from running the respective protocol on the best-performing set of physical qubits (using qubits 58, 59, 60, 61). In this case, RNXY4+PS yields $\mathcal{F}\approx 92\%$ after $55\mu$s (for both cases), significantly higher than the fidelity using postselection after error detection alone. Also shown is the best physical Bell state prepared on the same set of qubits (yellow squares), which, without DD (Best Physical No DD; pink) performs comparably to No DD+PS, though with much larger crosstalk-induced fidelity oscillations.  Applying staggered XY4 (Best Physical SXY4; yellow) removes the crosstalk and improves the best physical Bell state fidelity beyond RNXY4's (decaying to $\mathcal{F}\approx 71\%$ after $55\mu$s), but remains well below the performance of RNXY4+PS.}
\label{fig:res-3}
\end{figure*}
%%%%%%%%%%%%%%%%%%%%%%%%%%%%%%%%

\subsubsection{Logical $X$ and $Y$ errors}

So far, we have only discussed the detection and suppression of $Z$-type logical errors. In order to detect $X$ and $Y$-type logical errors, we proceed in analogy to the $\overline{Z}$ error detection procedure, but with a different unencoding step. Namely, we start by encoding the logical $\ket{\Phi_+}$ state, and then unencode in either $\ket{\Psi_+}$ to detect $X$-type errors or in $\ket{\Psi_-}$ to detect $Y$-type errors. The results are shown in \cref{fig:res-2} where, for comparison, we have also included the no-DD $\ket{\Phi_+}\mapsto\ket{\Phi_-}$ and $\ket{\Psi_+}\mapsto\ket{\Psi_-}$ results shown in \cref{fig:res-1}(b,d), which measure $Z$-type logical errors. 
It is clear from \cref{fig:res-2} that while unsuppressed logical errors $Z$ accumulate rapidly, logical errors of $X$ and $Y$-type grow much more slowly. This confirms that $ZZ$ crosstalk is the main source of logical errors. 

\subsubsection{NDD with postselection}

Having implemented NDD, we can further improve the results by using the error detection capability of the $[[4,2,2]]$ code, which allows us to perform postselection.
We do so by discarding any measurement outcome outside the logical basis (which would be the result of a physical error); i.e., we only keep measurement results corresponding to the bitstrings $\{0000,0110,1001,1111\}$. Doing so leads to our central experimental result, the \emph{demonstration of high-fidelity entangled logical qubits}, shown in \cref{fig:res-3}.

In \cref{fig:res-3}(a) and (b) we show the fidelities of $\ket{\Phi_+}\mapsto\ket{\Phi_-}$ and $\ket{\Psi_+}\mapsto\ket{\Psi_-}$ for dataset 2 (\cref{sec:expt-design}). Without DD, the fidelity decays rapidly while exhibiting crosstalk oscillations, as seen for dataset 1 in \cref{fig:res-1}. The fidelity improves once we perform postselection as described above. This leaves us with bitstrings corresponding to logical states, but logical errors still reduce the fidelity. Using NDD in the form of the RNXY4 sequence -- which suppresses both logical errors and physical errors -- followed by postselection, we achieve fidelities $>82\%$ for $\ket{\Phi_+}$ and $>87\%$ for $\ket{\Psi_+}$ over a $55\mu$s period. The average fidelities over the same period are 91.12$\%$ for $\ket{\Phi_+}$ and 93.66 $\%$ for $\ket{\Psi_+}$. For the particular set of qubits numbered $\{58, 59, 60, 61\}$ (light dashed lines), we find that the combination of NDD and postselection yields average fidelities of $\approx 95.44\%$ for $\ket{\Phi_+}$ and $\approx 94.78\%$ for $\ket{\Psi_+}$, which is significantly higher than when we use only the error detection capability of the code: $\approx 44.81\%$ for $\ket{\Phi_+}$ and $\approx 48.47\%$ for $\ket{\Psi_+}$.

\subsubsection{Beyond breakeven and the state-of-the-art}

\cref{fig:res-3} also includes results for physical (unencoded) Bell states. Here we show only the best Bell pair among all pairs we tested, both without DD and with staggered (crosstalk-robust~\cite{Zhou2023PRL}) XY4. The former (`Best Physical No DD') exhibits strong crosstalk-induced oscillations with an overall fidelity comparable to that of the mean logical encoded Bell pairs case without DD (No DD+PS). Adding SXY4 significantly improves the fidelity and outperforms even the mean fidelity of logical Bell pairs with NDD (RNXY4). This shows that NDD by itself is not better than working with physical qubits and a crosstalk-robust DD sequence. However, physical Bell pairs with SXY4 are significantly worse than RNXY4+PS, i.e., the case of NDD with postselection on the results of the $[[4,2,2]]$ code. \emph{This constitutes clear evidence of beyond-breakeven performance for our QEC-NDD strategy}.

Overall, it is clear that the combination of NDD and postselection significantly boosts the fidelity of logical Bell states. Moreover, our results improve upon the current state-of-the-art using superconducting transmon qubits.
E.g., Ref.~\cite{zhang2024demonstratinguniversallogicalgate} used distance $d=2$ surface codes 
to encode the four logical Bell states, with a peak encoding fidelity of 79.5$\%$. In contrast, we find an average postselected encoding fidelity of 98.05$\%$. The averaging is over the $10$ sets of qubits and over the two logical Bell states we prepare. 

Ref.~\cite{Hetenyi2024PRXQuantum} used the heavy-hex surface code with variable distance $d$ and reported a peak postselected fidelity of 93.7$\%$ after the first stabilizer round for $d=2$, declining to $\approx 30\%$ after five rounds, which corresponds to $\approx 27\mu$s on \texttt{ibm\_torino}.
In contrast, we find an average peak postselected encoding fidelity of 98.05$\%$ that declines to 84.87$\%$ after $55\mu$s, and a peak postselected fidelity of 98.00$\%$ (also averaged over the two logical Bell states we prepare) for the best set of qubits, that declines to 92.89$\%$ after $55\mu$s.

%%%%%%%%%%%%%%%%%%%%%%%%%%%%%%%%%%%%
\begin{figure}[t]
\hspace{0cm}{\includegraphics[scale=.55]{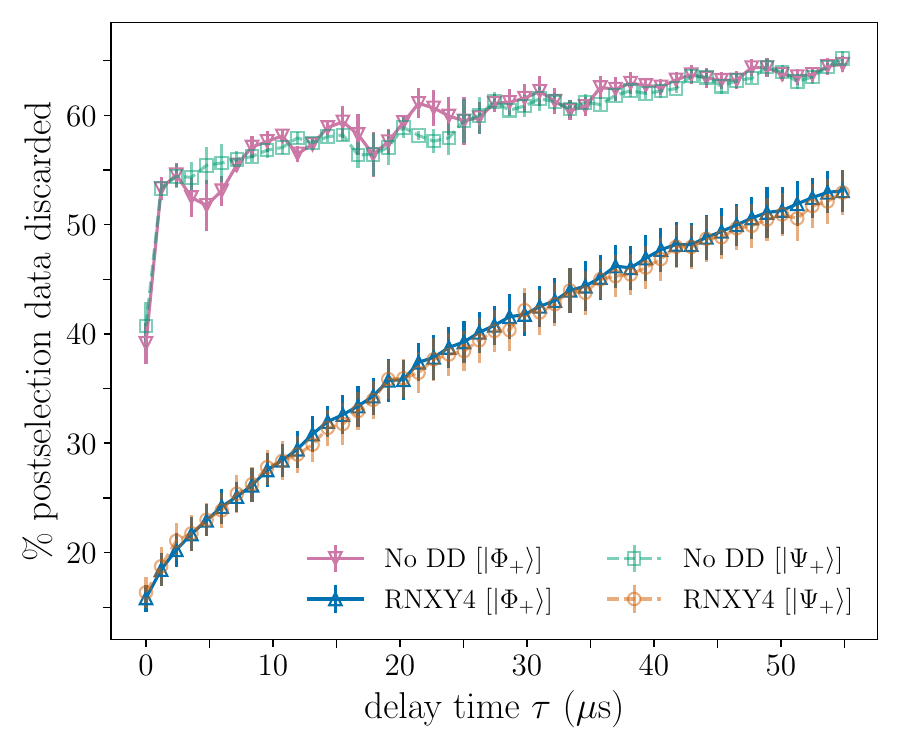}} 
\caption{The percentage of data discarded per circuit as a function of the delay time between encoding and unencoding (dataset 2). While NDD suppresses logical errors, it also reduces certain physical errors, as evidenced by the lower data discard rate when using NDD (i.e., fewer physical errors are detected).}
\label{fig:res-4}
\end{figure}
%%%%%%%%%%%%%%%%%%%%%%%%%%%%%%%%

\subsection{Physical error suppression and detection by NDD}

As explained in the discussion following \cref{lem:2} and in \cref{sec:illustr}, the NDD sequences suppress not only logical errors but physical errors as well. To see this, consider, e.g., the RNXX sequence [\cref{fig:schematic}(d)]. As described in \cref{sec:logZerr}, reading the pulse sequence vertically, each time step of this sequence operates in the logical subspace as $\overline{XX}$. Simultaneously, reading the sequence horizontally over a complete round of NDD, all four physical qubits undergo the physical XX sequence (i.e., $X-\tau-X-\tau$), which suppresses the set of physical errors $\{Y_i, Z_i\}_{i=1}^4$. Similarly, for the RNXY4 sequence [\cref{fig:schematic}(e)], each physical qubit undergoes UR$_4$. which robustly~\cite{Genov2017PRL} suppresses the sets of physical errors $\{Y_i, Z_i\}_{i=1,3}$ (due to the $X\tilde{X}\tilde{X}X$ sequence) and $\{X_i, Z_i\}_{i=2,4}$ (due to the $Y\tilde{Y}\tilde{Y}Y$ sequence).

One way to gauge the impact of this suppression of physical errors is shown in \cref{fig:res-4}, which displays the percentage of discarded data per circuit as a result of postselection over $55\mu$s (out of $4000$ shots), comparing No DD to RNXY4 for the two logical Bell states $\ket{\Phi_+}$ and $\ket{\Psi_+}$. The results are obtained through bootstrapping across the ten sets of four qubits used in these experiments. Significantly less data is discarded with RNXY4 than without DD. Since discarded data corresponds to the detection of physical errors, this means that NDD not only enhances the fidelity of logical Bell states by suppressing logical errors, but also reduces the occurrence of physical errors. 

\begin{figure}[t]
\hspace{0cm}{\includegraphics[scale=.55]{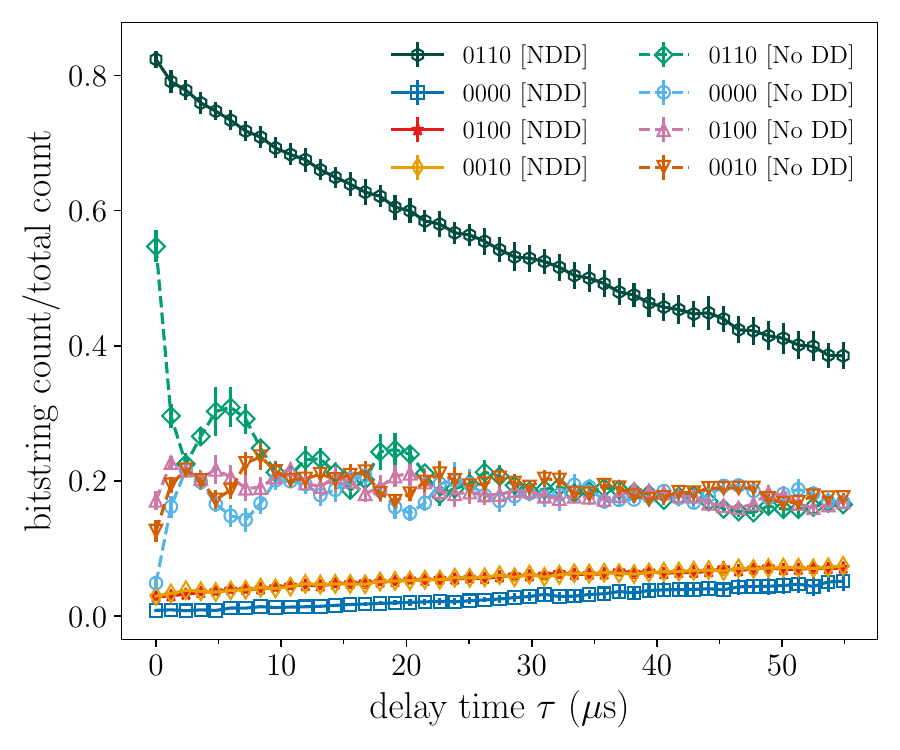}} 
\caption{The relative count of the $\{0000,0110,0100,0010\}$ bitstrings in the $\ket{\Phi_+}\mapsto \ket{\Phi_-}$ experiments (dataset 2). The two even-weight bitstrings correspond to the logical fidelity ($0110$) and logical $Z$ error ($0000$), and the two odd-weight bitstrings correspond to the detection of physical $\{Z_i\}_{i=1}^4$ errors. Dashed (solid) lines correspond to no DD (RNXY4). The NDD sequence, in addition to suppressing the logical errors and improving the logical fidelity, also strongly suppresses the occurrence of these physical errors.}
\label{fig:res-5}
\end{figure}

We can go further and use Algorithmic Error Tomography~\cite{Pokharel2024npj} to identify specific physical error types. For example, the $0100$ and $0010$ bitstrings correspond, respectively, to physical $Z_2, Z_4$ and $Z_1, Z_3,$ errors in the $\ket{\Phi_+}\mapsto \ket{\Phi_-}$ logical Bell state experiment. In \cref{fig:res-5} we show the corresponding relative bitstring counts, which are measures of detecting either of these errors. Additionally, we show the $0110$ and $0000$ relative bitstring counts, which correspond to the logical fidelity and the logical $Z$ error, respectively. Without DD (dashed), we observe that the logical fidelity is low, and both logical and physical errors increase and oscillate. With DD (solid), the RNXY4 sequence strongly suppresses both types of errors.  

These results demonstrate that, as claimed, NDD sequences suppress \emph{both} physical and logical errors. 

%%%%%%%%%%%%%%%%%%%%%%%%%%%%%%%%%%%%
%%%%%%%%%%%%%%%%%%%%%%%%%%%%%%%%

%%%%%%%%%%%%%%%%%%%%%%%%%%%%
%%%%%%%%%%%%%%%%%%%%%%%%%%%%
%%%%%%%%%%%%%%%%%%%%%%%%%%%%

\section{Discussion}
\label{sec:discussion}

The operation of QEC codes is adversely affected by the occurrence of logical errors that the code cannot detect or correct. Here, we have shown how to combine QEC with dynamical decoupling implemented in terms of the 
%logical operators 
{normalizer elements} of the code, resulting in a hybrid QEC-{N}DD strategy that is significantly more effective than either QEC or {N}DD alone. 
We designed our NDD sequences to simultaneously perform logical error suppression and to be robust DD sequences at the physical level, resistant to both control errors and crosstalk. As an added benefit, these sequences suppress many physical errors as well. Our results using the $[[4,2,2]]$ code and IBM transmon qubits, featured in \cref{fig:res-3}, demonstrate a beyond-breakeven fidelity of entangled logical qubits. The fidelities we report are the highest to date for entangled logical qubits using superconducting qubits.

Our findings address a need along the path toward fault-tolerant quantum computation: keeping codes relatively small and nimble while still effectively handling logical errors. 
Future research should aim to optimize NDD sequences tailored to specific codes and integrate QEC-NDD into quantum algorithms. 
Another interesting future direction is the optimization of QEC-NDD for tunable-coupler transmon devices; we present preliminary results in \cref{app:marrakesh}.

\begin{acknowledgments}
This material is based upon work supported by, or in part by, the Intelligence Advanced Research Projects Activity (IARPA), under the Entangled Logical Qubits program through Cooperative Agreement Number W911NF23-2-0216, by the U.S. Army Research Laboratory and the U.S. Army Research Office under contract/grant number W911NF2310255, and by the Defense Advanced Research Projects Agency under Agreement HR00112230006. The views, opinions and/or findings expressed are those of the author(s) and should not be interpreted as representing the official views or policies of the Department of Defense or the U.S. Government. This research was conducted using IBM Quantum Systems provided through the University of Southern California's IBM Quantum Innovation Center. The views expressed are those of the authors and do not reflect the official policy or position of IBM or the IBM Quantum team.
\end{acknowledgments}

%%%%%%%%%%%%%%%%%%%%%%%%%%%%
%%%%%%%%%%%%%%%%%%%%%%%%%%%%
%%%%%%%%%%%%%%%%%%%%%%%%%%%%
\appendix

\section{Quantum codes and logical errors}
\label{app:QECC}

A general $[[n,k,d]]$ code encodes $k$ logical qubits into $n$ physical qubits with distance $d$~\cite{Knill:1997kx}.
The weight $w$ of an error is the number of physical qubits it affects simultaneously. An $[[n,k,d]]$ stabilizer code corrects every error whose weight satisfies $w \le t = \bigl\lfloor \tfrac{d-1}{2}\bigr\rfloor$
and detects every error whose weight $w \le d-1$. A $d=2$ code is a pure quantum error detection (QED) code, and a distance-$d$ QEC code can always be used as a QED code for errors of weight $<d$. The code distance is the minimum number of physical qubits that must experience an error to cause an undetectable logical error, i.e., an error forming a logical operation inside the code space. These logical errors can either be inherently present or result from the accumulation of lower-weight errors over time. For example, in the context of superconducting qubits with fixed-frequency couplers, a prevalent challenge is $ZZ$ crosstalk~\cite{Ku2020,Wei:2022aa}, which inherently introduces weight-two errors that can present as logical errors for distance-$2$ codes. 

An $[[n,k,d]]$ stabilizer code $\mC$ is defined as the $+1$ eigenspace of a stabilizer group $\mS$ of order $2^{n-k}$ (with commuting generators $\{S_j\}_{j=1}^{n-k}$).
A stabilizer group is any subgroup of $\mP_n$ that excludes $-I$ (which implies that it is Abelian).
One can factor the $n$-qubit Hilbert space into $k$ logical qubits with an associated group of canonical logical operators $\mL  = \langle i, \overline{X}_j,\overline{Z}_j\rangle_{j=1}^k \subset\mP_n$ (we use $\langle \cdots \rangle$ to denote a generating set; $\abs{\mL } = 4^{k+1}$) and $n-k$ syndrome qubits (which can be used to detect errors). Specifically, syndrome qubits can be used to detect Pauli group terms that anticommute with at least one of the $n-k$ stabilizer generators~\cite{Gottesman1996PRA}.
Since $\mS$ is commutative, operators in $\mS$ can be simultaneously diagonalized. This simultaneous diagonalization partitions the Hilbert space into an orthogonal sum of $2^{n-k}$ subspaces {(known as syndrome spaces)}, each of dimension $2^k$, corresponding to $2^{n-k}$ choices of {$\pm 1$} eigenvalues of $S_j$ for $j = 1, \dots, n-k$, 
i.e., different values of the syndrome. 
$\mC$ is associated with the trivial (no error, corresponding to $+1$ eigenvalue of all $S_j$) syndrome, and two logical operators act identically on $\mC$ if they only differ by a stabilizer element, i.e., the full group of logical operators
is $\NS = \mS  \mL$, the {normalizer} of $\mS$ in $\mP_n$.

For a code $\mC  \subset (\mathbb{C}^2)^{\otimes n}$, the stabilizer group $\mS$ is uniquely defined as  
\begin{equation}
  \mS  = \{P \in \mP_n: \forall \ket{\psi} \in \mC , P \ket{\psi} = \ket{\psi}\}.
\end{equation}
The normalizer (or centralizer) of $\mS$ in $\mP_n$ is the group $\NS$ of Pauli operators commuting with all elements of $\mS$.%
\footnote{{The normalizer $\NS = \mathcal{N}_{\mG}(\mS )$ of $\mS  \subseteq \mG $ is defined as $\{g \in \mG  \text{ s.t. } g\mS g^{-1} = \mS \}$, and the centralizer $\mathcal{Z}(\mS ) = \mathcal{Z}_{\mG}(\mS )$ of $\mS  \subseteq \mG $ is defined as $\{g \in \mG  \text{ s.t. } \forall s \in \mS , gs = sg\}$. These two notions coincide in the case of stabilizer groups.}}
{The normalizer} is also uniquely defined. 
However, the group of canonical logical operators is not unique: one can choose arbitrary stabilizers 
and multiply the canonical generators $\bar{X}_j, \bar{Z}_j$ by those to obtain other choices of the group $\mL$.

Another important set is 
$\mD = \mP_n \setminus \NS$, the set of detectable errors. Since every pair of elements of $\mP_n$ either commute or anticommute, and $\NS$ contains all elements of $\mP_n$ that commute with $\mS$, every element of $\mD$ must anticommute with at least one element of $\mS$.

\section{[[4,2,2]] code}
\label{app:422}

The $[[4,2,2]]$ code is an error detection code that encodes $k=2$ logical qubits into $n=4$ physical qubits~\cite{Vaidman:1996vs}. The stabilizer group is $\mS =\{ I, XXXX, YYYY, ZZZZ\}$. Defining the logical states as 
\begin{equation}
    \begin{aligned}
& \ket{\overline{00}} =\frac{\ket{0000}+\ket{1111}}{\sqrt{2}},\ \ 
\ket{\overline{10}} =\frac{\ket{0110}+\ket{1001}}{\sqrt{2}}, \\
& \ket{\overline{01}} =\frac{\ket{0011}+\ket{1100}}{\sqrt{2}},\ \ \ket{\overline{11}} =\frac{\ket{1010}+\ket{0101}}{\sqrt{2}},
\end{aligned}
\end{equation}
a set of logical operators for the code can be defined such that $\overline{X}_1=XIIX$, and $\overline{X}_2=IIXX$, up to multiplication by a stabilizer element. Therefore, we have $\overline{XX}=X I X I$. Similar definitions apply to the logical $Z$ operators: $\overline{Z}_1=IIZZ$ and $\overline{Z}_2=ZIIZ$, allowing us to form the full logical Pauli group. Using the same definitions, we have $\overline{\text{CNOT}}_{12}=\text{SWAP}_{12}$ and similarly, $\overline{\text{CNOT}}_{21}=\text{SWAP}_{23}$, i.e., we can perform logical CNOTs by swapping the physical qubits. 

Our particular choice of logical operators is motivated by the fact that $\overline{XX}$ and $\overline{YY}$ (i.e., the logical operators we use to implement the LDD or NDD sequences) have a natural staggering of their physical $X$ and $Y$ gates. In other words, in the implementation of the logical operators comprising NDD, nearest-neighbor qubits are always interleaved with an identity operation (e.g., $IXIX$ as opposed to $IXXI$). This is critical for nearest-neighbor crosstalk cancellation~\cite{Zhou2023PRL}.

%%%%%%%%%%%%%%%%%%%%%%%%%%%%
%%%%%%%%%%%%%%%%%%%%%%%%%%%%

To encode the logical Bell states $\ket{\Psi_\pm}$ and $\ket{\Phi_\pm}$ using the two logical qubits of the $[[4,2,2]]$ code, consider:
\bes
\begin{align}
    \ket{\Phi_\pm} &= \frac{1}{\sqrt{2}}\big(\ket{\overline{00}}\pm\ket{\overline{11}}\big)\\
    &= \frac{1}{\sqrt{2}}\big(\ket{0000}+\ket{1111}\pm\ket{0101}\pm\ket{1010}\big) \\
    &= \overline{\text{CNOT}}_{21}\frac{1}{\sqrt{2}}\big[\ket{0000}+\ket{1111}\pm\ket{0011}\pm\ket{1100}\big] \\
    &= \overline{\text{CNOT}}_{21}\frac{1}{\sqrt{2}}\big[(\ket{00}\pm\ket{11})(\ket{00}\pm\ket{11})\big]. 
\end{align}
\ees
Thus, we proceed by preparing two \textit{physical} copies of the target Bell state on the four physical qubits, then apply $\overline{\text{CNOT}}_{21}$, which creates the intended Bell state. 
The $\ket{\Psi_\pm}$ Bell states are prepared similarly.

\section{Construction of our $[[4,2,2]]$ code subgroup-NDD sequence} \label{app:full-LDD}

One of the limitations of our experimental setup is that physical $Z$ gates are unavailable: fixed-frequency transmons utilize virtual-$Z$ gates~\cite{McKay:2017tv,Vezvaee2024arxiv}. Such gates are problematic for DD~\cite{Vezvaee2024arxiv}. Here, now demonstrate that there are DD sequences satisfying the conditions of \cref{lem:2} which do not use any physical $Z$ pulses in the context of the $[[4,2,2]]$ code.

To construct such a sequence, we first choose generators of $\mL$ that do not involve physical $Z$'s. A simple example is $\mL  = \langle i, XIIX, IIXX, IIYY, YIIY \rangle$. One could follow the construction below with this choice. However, in order to improve the ability of the DD sequence to suppress $ZZ$ crosstalk, we make a different choice.
First, multiply the original generators of $\mL$ (that is, $XIIX, IIXX, IIZZ, ZIIZ$) by the stabilizers $XXXX$, $XXXX$, $XXXX$, and $ZZZZ$, respectively, to obtain $\mL ' = \langle i, IXXI, XXII, XXYY, IZZI \rangle$. Then, after some group operations,  we can write the same group using a different set of generators containing no $Z$'s: $\mL ' = \langle i, XXII, IIYY, IYIY, XIXI \rangle$. The {corresponding N}DD sequence consisting of two repetitions of $XIXI$, $IYIY$, $XIXI$, $IIYY$, $XIXI$, $IYIY$, $XIXI$, and $XXII$, which implements
the decoupling group $\mL ' / \{\pm 1, \pm i\}$, and involves no $Z$ gates. 

This sequence was obtained as follows: denote the generators of $\mL$ (or $\mL '$) as $h_1, h_2, h_3, h_4$ and choose any Gray code---%(aka reflected binary code)
a sequence of bitstrings $a_0, \dots, a_{15}$ such that neighboring bitstrings differ in only a single digit. Then, pick $g_j = h^{a_j}$ (for $j=0,\dots, 15$), where $a_j$ is interpreted as a multiindex (i.e., $g_j = \prod_{k=1}^{4} h_k^{(a_j)_k}$). The pulses of the DD sequence are $g_jg_{j+1}$ for $j = 0, \dots, 15$.

%%%%%%%%%%%%%%%%%%%%%%%%%%%%
\section{Issues involving the implementation of logical state tomography using the $[[4,2,2]]$ code}
\label{app:tomo}

A general two-qubit state requires information about $15$ expectation values which requires $9$ independent measurement settings (achieved by measuring each qubit in three complementary bases; e.g., $X$, $Y$, and $Z$)~\cite{james:052312}. This holds for logical tomography of a logical two-qubit state as well. There are a few ways in which a logical observable can be measured. First, one can use stabilizer-measurement-like circuits where the logical operator is measured by executing a circuit composed of CNOT gates targeting an ancilla qubit that is measured to learn the logical measurement outcome. This method can further be paired with a round of syndrome extraction using additional ancilla qubits to learn whether the state was in the code space to begin with. This combination of measurements allows us to invoke the code's protection while simultaneously performing a measurement. The downside of this method is the additional overhead in CNOTs and ancilla qubits needed to perform the measurement protocol. In our case, this would require a substantial overhead in SWAP gates as well since the IBM QPU's heavy-hex lattice does not pair naturally with the $[[4, 2, 2]]$ code. Using this logical measurement method would inevitably introduce more errors and reduce the accuracy of logical state tomography. 

An alternative, less costly approach for performing logical measurements is to directly measure all data qubits of the code in lieu of introducing ancillas and additional CNOTs. However, this method is incompatible with extracting information about the entire stabilizer generator set while also performing the logical measurement. As a result, we would not be able to know with certainty that the system state was in the code space at the time of measurement.  There is some nuance to this approach, which does allow us to learn information about stabilizers which commute qubit-wise (i.e., they share the same Pauli operator or $I$ on the same qubit) with the logical operator being measured.  For example, we can simultaneously measure $\overline{X}_1=XIIX$,  $\overline{X}_2=IIXX$, $\overline{X_1X_2}=X I X I$ and the stabilizer $S_1 = XXXX$ since these operators all qubit-wise commute. Performing the measurement of each data qubit in the $X$ basis gives us information about each of these logical observables, as well as information about whether or not the system was in the logical code space with respect to the stabilizer generator $S_1$. Unfortunately, this symmetry does not hold for the operators $\overline{Y}_1 = XIZY$, $\overline{Y}_2 = ZIXY$, and $S_2=YYYY$ since these operators do not all commute qubit-wise. This means that we cannot simultaneously learn information about these logical observables and information about whether or not the system was in the code space with respect to the stabilizer $S_2$ in the same shot, reducing our total information about the system. This would lead to a less precise estimation of the logical density matrix, as we will count more experimental shots involving the system outside of the logical state space in our estimation of the logical expectation value. 

For these reasons, we did not use logical state tomography in this work.

%%%%%%%%%%%%%%%%%%%%%%%%%%%%
\section{Creating robust logical sequences}
\label{app:cross}

As discussed in the main text, we use the logical operators $\overline{XX}=XIXI$ and $\overline{YY}=IYIY\equiv YIYI$ (along with identity operators) to generate the LDD groups ${\mG}_{{\text{LXY4}}}=\{I,\overline{XX},\overline{YY}, \overline{ZZ}\}$, and ${\mG}_{{\text{LXX}}}=\{I,\overline{XX}\}$. Cycling over the group elements creates the basis for the four-pulse sequences shown in the red boxes of \cref{fig:schematic}(d,e). However, we add two additional components to enhance crosstalk cancellation and robustness.

Since the main source of logical errors is $ZZ$ crosstalk, we modify LDD so that it cancels such errors. Thus, in the case of LXX, we insert a stabilizer every other pulse-step to consistently apply a staggered sequence to all qubits; in this manner, we move from LDD to NDD. These sequences correspond to applying logical $X$ and $Y$ operators to both logical qubits.

To create the robust sequences, we mirror the original sequences (hence the eight-pulse sequences) and instead of $X$ and $Y$, we apply $\tilde{X} = R_x(-\pi)$, $\tilde{Y}= R_y(-\pi)$, respectively, such that each sequence undergoes a robust pulse sequence at the physical level. We also apply staggered physical XY4 to each qubit as shown in \cref{fig:schematic}(f). This sequence is also inherently robust at the physical level since each qubit receives an XY4 sequence, which is robust to pulse errors~\cite{Genov2017PRL}. However, it is not a universal decoupling sequence at the logical level, as seen in the figure (i.e., it lacks the ability to decouple arbitrary single logical-qubit errors).

%%%%%%%%%%%%%%%%%%%%%%%%%%%%%%%%%%%%
\begin{figure*}[t]
\hspace{0cm}{\includegraphics[scale=.65]{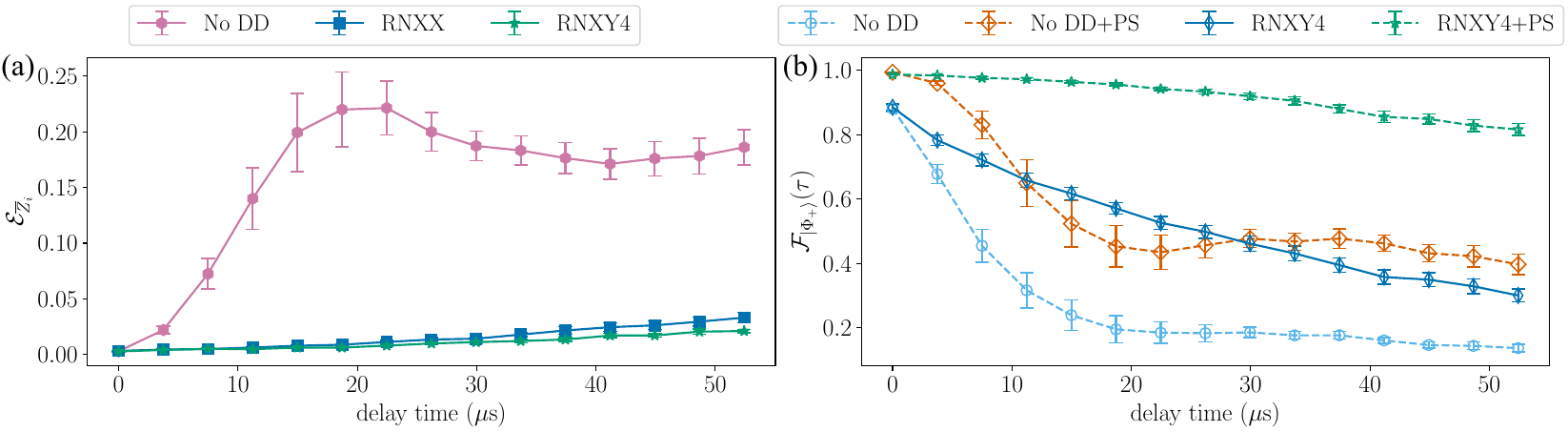}}
\caption{Performance of NDD on for an encoded $\ket{\Phi_+}$ Bell state on \texttt{ibm\_marrakesh} for a single set of qubits. (a) shows the logical errors and (b) shows the postselection results. This device features tunable couplers that significantly reduce the crosstalk. Nevertheless, accumulation of dephasing errors over time (up to $\approx 55~\mu$s) leads to the appearance of logical errors. The same NDD sequence we used in the main text successfully suppresses these errors, as shown in (a). Additionally, NDD with postselection outperforms postselection alone, as shown in (b).}
\label{fig:marrakesh}
\end{figure*}
%%%%%%%%%%%%%%%%%%%%%%%%%%%%%%%%

\section{Normalizer dynamical decoupling with tunable couplers}
\label{app:marrakesh}

Here we present additional experiments on \texttt{ibm\_marrakesh}, which features tunable couplers~\cite{StehlikPRL2021} unlike the always-on $ZZ$ interaction in \texttt{ibm\_kyiv}. Consequently, for this set of experiments, crosstalk is significantly reduced compared to the other set of results ($\lesssim  5$ kHz vs tens of kHz). Nevertheless, a combination of dephasing and residual crosstalk accumulates, leading to logical errors. \Cref{fig:marrakesh} illustrates the performance of NDD sequences on this QPU, averaged over $9$ different qubit sets. \Cref{fig:marrakesh}(a) [equivalent to \cref{fig:res-1}(b)] shows that without DD, logical errors remain unsuppressed. However, both NDD sequences we employ successfully suppress these errors. \Cref{fig:marrakesh}(b) [equivalent to \cref{fig:res-3}(a)] demonstrates that using only postselection, the fidelity averages to $\mathcal{F} \approx 56.23\%$, whereas combining postselection with NDD increases it to $\mathcal{F} \approx 91.73\%$. 

Optimizing NDD for specific types of tunable couplers to maximize the interplay between DD and couplers is an avenue for future work.

%%%%%%%%%%%%%%%%%%%%%%%%%%%%
%%%%%%%%%%%%%%%%%%%%%%%%%%%%
\bibliography{biblo}

\end{document}